\documentstyle{article}

\def\d{\partial}
\def\p{\partial\over\partial}

\def\bea{\begin{eqnarray}}
\def\eea{\end{eqnarray}}
\def\nn{\nonumber}

\def\beq{\begin{equation}}
\def\eeq{\end{equation}}
\def\ba{\beq\new\begin{array}{c}}
\def\ea{\end{array}\eeq}
\def\be{\ba}
\def\ee{\ea}
\def\stackreb#1#2{\mathrel{\mathop{#2}\limits_{#1}}}
\def\Tr{{\rm Tr}}
\def\res{\hbox{res}}
\def\f{1\over}
\def\r{{r\over 2}}
\makeatletter
\newdimen\normalarrayskip              
\newdimen\minarrayskip                 
\normalarrayskip\baselineskip
\minarrayskip\jot
\newif\ifold             \oldtrue            \def\new{\oldfalse}
\def\arraymode{\ifold\relax\else\displaystyle\fi} 
\def\eqnumphantom{\phantom{(\theequation)}}     
\def\@arrayskip{\ifold\baselineskip\z@\lineskip\z@
     \else
     \baselineskip\minarrayskip\lineskip2\minarrayskip\fi}
\def\@arrayclassz{\ifcase \@lastchclass \@acolampacol \or
\@ampacol \or \or \or \@addamp \or
   \@acolampacol \or \@firstampfalse \@acol \fi
\edef\@preamble{\@preamble
  \ifcase \@chnum
     \hfil$\relax\arraymode\@sharp$\hfil
     \or $\relax\arraymode\@sharp$\hfil
     \or \hfil$\relax\arraymode\@sharp$\fi}}
\def\@array[#1]#2{\setbox\@arstrutbox=\hbox{\vrule
     height\arraystretch \ht\strutbox
     depth\arraystretch \dp\strutbox
     width\z@}\@mkpream{#2}\edef\@preamble{\halign
\noexpand\@halignto
\bgroup \tabskip\z@ \@arstrut \@preamble \tabskip\z@ \cr}%
\let\@startpbox\@@startpbox \let\@endpbox\@@endpbox
  \if #1t\vtop \else \if#1b\vbox \else \vcenter \fi\fi
  \bgroup \let\par\relax
  \let\@sharp##\let\protect\relax
  \@arrayskip\@preamble}
\def\eqnarray{\stepcounter{equation}%
              \let\@currentlabel=\theequation
              \global\@eqnswtrue
              \global\@eqcnt\z@
              \tabskip\@centering
              \let\\=\@eqncr
              $$%
 \halign to \displaywidth\bgroup
    \eqnumphantom\@eqnsel\hskip\@centering
    $\displaystyle \tabskip\z@ {##}$%
    \global\@eqcnt\@ne \hskip 2\arraycolsep
         $\displaystyle\arraymode{##}$\hfil
    \global\@eqcnt\tw@ \hskip 2\arraycolsep
         $\displaystyle\tabskip\z@{##}$\hfil
         \tabskip\@centering
    &{##}\tabskip\z@\cr}
\def\theequation{\thesection.\arabic{equation}}
\newfont{\hr}{msbm10}
\newfont{\ams}{msam10}

\begin{document}
\begin{titlepage}
\setcounter{footnote}0
\begin{center}
\hfill FIAN/TD-15/96\\
\hfill ITEP/TH-46/96\\

\vspace{0.3in}
{\LARGE\bf  More Evidence for the WDVV Equations in N=2 SUSY
Yang-Mills Theories}
\\
\bigskip\bigskip\bigskip

{\Large A.Marshakov}
\footnote{E-mail address:
mars@lpi.ac.ru, andrei@heron.itep.ru, marshakov@nbivax.nbi.dk},
{\Large A.Mironov}
\footnote{E-mail address:
mironov@lpi.ac.ru, mironov@heron.itep.ru}\\
{\it Theory Department, P. N. Lebedev Physics
Institute, Leninsky prospect 53, Moscow, ~117924, Russia\\
and ITEP, Moscow ~117259, Russia}\\and\\
{\Large A.Morozov}
\footnote{E-mail address:
morozov@vxdesy.desy.de}
\\
{\it ITEP, Moscow ~117 259, Russia}\\
\end{center}
\bigskip \bigskip

\begin{abstract}
We consider $4d$ and $5d$ ${\cal N}=2$ supersymmetric theories
and demonstrate that in general their Seiberg-Witten
prepotentials satisfy the Witten-Dijkgraaf-Verlinde-Verlinde
(WDVV) equations.
General proof for the Yang-Mills models (with matter in
the first fundamental representation) makes use of the
hyperelliptic curves and underlying integrable systems.
A wide class of examples is discussed, it contains few
understandable exceptions. In particular, in perturbative regime
of $5d$ theories in addition to naive field theory expectations
some extra terms appear, like it happens in heterotic string models.
We consider also the example of the Yang-Mills theory with
matter hypermultiplet in the adjoint representation
(related to the elliptic Calogero-Moser system)
when the standard WDVV equations do not hold.
\end{abstract}

\end{titlepage}

\newpage
\setcounter{footnote}0

\section{Introduction}

More than two years ago N.Seiberg and E.Witten proposed a way
\cite{SW1,SW2} leading to  considerable progress in
understanding of the low-energy effective actions.
Among other things, it provided a non-trivial confirmation
of the old {\it belief} that the low-energy effective actions
(the end-points of the renormalization group flows) fit into universality
classes depending on the vacuum structure of the theory.
The trivial (and generic) case is that of isolated vacua,
i.e. when there are no marginal degrees of freedom and effective
theories on the moduli space of the vacua. If degenerated vacua form
finite-\-dimensional varieties, the effective actions can be essentially
described in terms of $1d$ systems
\footnote{By $1d$ $(0+1)$ systems here we mean the conventional
classical/quantum mechanical systems with {\it finite}-dimensional phase
space (\# of degrees of freedom is equal to the rank of the gauge group).}
(representatives of the
same universality class), though the original
theory lives in a many-dimensional space-time. Moreover, these effective
theories can be integrable
\footnote{When degenerated vacua form
infinite-dimensional structures one should expect generalized integrability
(with spectral curve substituted by spectral hypersurface).}.
Originally,
this phenomenon was studied in the context of $2d$ topological theories,
while the Seiberg-Witten (SW) construction can be considered as
an extension of this approach from two to four space-time
(``world sheet'') dimensions.
The natural context where one encounters continuous degeneration of vacua in
four dimensions is {\it supersymmetry}; the SW construction {\it per se}
concerns the $4d$ ${\cal N}=2$ SUSY Yang-Mills models.  Integrable structure
behind the SW theory has been found in \cite{GKMMM} and later examined in
detail in \cite{SWI1}-\cite{N}. Its intimate relation to the previously
known topological theories has been revealed recently in \cite{MMM}
where it was proven that the $4d$ prepotentials do satisfy the
WDVV equations \cite{W,DVV}.
The purpose of the present paper is to give more evidence
(and more examples) confirming this relation.

Naively, the SW low-energy effective action is not
topological: it describes propagating particles. This is,
however, a necessary consequence of the vacuum degeneracy and
unbroken supersymmetry. All the correlators in the low-energy
theory can be expressed through the correlators of ``holomorphic''
(or ``pure topological'') subsector. Technically, propagators
are contained in the $\langle\phi\bar\phi\rangle$ correlators,
(derivatives of) which are related by the ${\cal N}=2$ supersymmetry
to the $\langle\phi\phi\rangle$ ones. This is summarized in the statement
(sometimes called special geometry) that the low-energy effective action is
{\it completely} expressed in terms of a {\it holomorphic} prepotential
$F(a_I)$, $\frac{\partial}{\partial\bar a_I} F = 0$.  Even if one accepts the
term ``topological'' for this kind of models, what we consider is a $4d$
topological theory, {\it a priori} very different from the familiar ($2d$)
examples. However, if one believes in the above-mentioned universality, it is
clear that only the structure of the moduli space of vacua (but not the
space-time dimension) is essential.  Since the vacuum variety in the SW
theory is finite-dimensional, one should expect that this theory is not too
far from the conventional topological models, devised to describe
exactly this kind of vacua; as was shown in \cite{MMM} and confirmed below in
the present paper this is indeed the case. The way to see this is to recall
that the prepotential of conventional topological theory satisfies the
WDVV equations \cite{W,DVV} and so does the prepotential of the SW theory
\cite{MMM} (similar claims were made in \cite{BM,dBdW}).

Mathematically, the WDVV equations reflect the specific properties of
the moduli space of Riemann spheres with punctures \cite{Ma,KoMa}, in the $2d$
context these spheres being interpreted as the world sheets.
Naturally, in the $4d$ context they could be substituted by
the four-dimensional world hypersurfaces, and naively instead of integrable
systems, associated with the spectral curves one would deal with
those, associated with spectral hypersurfaces -- with
no reason for the standard WDVV eqs. to emerge. This is of course
true for ``generic topological $4d$ models'', but this is not
the only relevant problem that can be addressed in four dimensions.
If one instead emphasizes the issue of vacuum structures and
universality classes, then the {\it standard} WDVV eqs should
be expected in the ({\it not} the most degenerate from the $4d$ point
of view) situation with  finite-dimensional (in the
space of all fields) variety of vacua -- but the conventional
derivation of these equations is not (at least, directly) applicable.
Moreover, as argued in \cite{MMM}, the SW example emphasizes
that the natural form (and thus the structure behind) the
WDVV eqs is covariant under the linear transformations of
distinguished (flat) coordinates on the moduli space of vacua --
what is true, but not {\it transparent} in the conventional
presentation. For us, all this implies that the true
origin of the WDVV eqs is still obscure.

Our goal in this paper is rather modest: given (relatively)
explicit expressions for particular SW prepotentials
we check that they indeed satisfy the WDVV eqs.

In sect.\ref{eqs} we briefly remind what are the
WDVV eqs and how they are related in \cite{MMM} to
the algebra of meromorphic 1-differentials on spectral curves.

Then in sect.\ref{pert} we discuss the perturbative prepotentials
in $4d$ and $5d$ ${\cal N}=2$ SUSY Yang-Mills theories which
are expected to satisfy the WDVV eqs themselves. We find that this is indeed
true, but only when matter hypermultiplets
are absent or belong to the first fundamental representation
of the gauge group (the representation of the lowest dimension). Remarkably,
this is in agreement with the
fact that no Seiberg-Witten curves are known beyond such cases,
and there are no string model to produce them in the field theory limit.

In sect.\ref{npsw} the WDVV eqs are derived at the non-perturbative
level from the representation
of the prepotential in terms of spectral Riemann surfaces.

The general reasoning is illustrated in sect.\ref{Examples} by examples
which include
${\cal N}=2$ SUSY YM models with classical simple gauge groups and the
matter hypermultiplets in the first fundamental representation.
Also the $5d$ $SU(N_c)$ pure gauge model is analyzed, and Seiberg-Witten
theory is shown to produce a prepotential, corrected as compared
to the naive field-theory expectations. Exactly these corrections are
necessary to make the WDVV equations true.

Finally, in sect.\ref{Calogero} the $SU(N_c)$ model with the matter
hypermultiplet in adjoint representation is analyzed in
detail. We demonstrate by explicit calculation that the WDVV equations
are {\it not} satisfied in this case. Technically the reason is that
the spectral curve is essentially non-hyperelliptic and the
algebra of 1-differentials (though existing and being closed) is
non-associative.
The deep reason for the breakdown of WDVV in its standard from
is the appearance of new modulus: the parameter $\tau$
of the complex structure on {\it bare} elliptic curve, which is interpreted
as the ultraviolet (UV) coupling constant of the UV-finite $4d$ theory.
Further understanding of the relevant deformation of the WDVV
equations (presumably related to the analogue of WDVV eqs for
the generating function of the elliptic Gromov-Witten classes
\cite{Getzler}) is important for further investigation of
string models (since $\tau$ is the remnant of heterotic dilaton
from that point of view).

\section{The full set of the WDVV equations \label{eqs}}
\setcounter{equation}{0}
The WDVV equations are non-linear relations for the third derivatives
of the prepotential
\be
F_{IJK} \equiv \frac{\partial^3 F}{\partial a_I
\partial a_J \partial a_K}.
\ee
written in the most convenient way in terms of matrices
$F_I$, $(F_I)_{MN} \equiv F_{IMN}$.
The {\it moduli} $a_I$ are defined up to linear transformations
(i.e. define the {\it flat structure} on the {\it moduli space})
which leave the whole set (\ref{WDVV}) invariant.

Any linear combination
of these matrices can be used to form a ``metric''
\be\label{metric}
Q = \sum_K q_KF_K.
\ee
Generically it is non-degenerate square matrix and can be
used to raise the indices:
\be\label{3}
C^{(Q)}_I \equiv Q^{-1}F_I, \ \ \ {\rm or}\ \
{C^{(Q)}}^M_{IN} = (Q^{-1})^{ML}F_{ILN}.
\ee
The WDVV equations state that for any given $Q$ all the
matrices $C^{(Q)}_I$ commute with each other:
\be
C^{(Q)}_IC^{(Q)}_J = C^{(Q)}_JC^{(Q)}_I \ \ \ \ \ \forall I,J
\label{WDVV}
\ee
In particular, if $Q = F_K$, then $C^{(K)}_I = F_K^{-1}F_I$ and \cite{MMM}
\be
F_IF_K^{-1}F_J = F_JF_K^{-1}F_I, \ \ \ \forall I,J,K.
\label{WDVVikj}
\ee
If (\ref{WDVVikj}) holds for some index $K$, say $K=0$,
it is automatically true for any other $K$
(with the only restriction for $F_K$ to be non-degenerate).
Indeed,
\footnote{
This simple proof was suggested to us by A.Rosly.
}
since $F_I = F_0 C_I^{(0)}$,
\be
F_IF_K^{-1}F_J =
F_0 \left(C_I^{(0)}(C_K^{(0)})^{-1} C_J^{(0)}\right)
\ee
is obviously symmetric w.r.t. the permutation $I\leftrightarrow J$.
Of course, it also holds for any other non-degenerate $Q$ in
(\ref{metric}).

Eqs.(\ref{WDVV}) are just the associativity condition of the formal algebra
\be
\varphi_I\circ \varphi_J = {C^{(Q)}}_{IJ}^K
\varphi_K, \nn \\
(\ref{WDVV}) \Leftrightarrow
(\varphi_I \circ \varphi_M )\circ \varphi_J =
\varphi_I \circ (\varphi_M \circ \varphi_J).
\label{alge}
\ee
The way to prove the WDVV eqs, used in \cite{MMM} and
in the present paper, is to obtain (\ref{alge}) from
a family of algebras formed by a certain set of (meromorphic) $(1,0)$-forms on
the curves, associated with particular
$4d$ Yang-Mills models:
\be
dW_I(\zeta) dW_J(\zeta) = {C^{(Q)}}_{IJ}^K dW_K(\zeta)
dQ(\zeta) \ {\rm mod}\ \left(d\omega(\zeta),d\lambda(\zeta)\right).
\label{algdiff} \label{2.8}
\ee
If this algebra exists and is associative we get the WDVV equations in context
of the SW approach.
The structure constants $C_{IJ}^K$ depend on the choice of the linear
combination $dQ(\zeta) = \sum_K q_KdW_K(\zeta)$.
In all relevant cases
$dW_I(\zeta) \sim P_I(\zeta)$, where for appropriately chosen
coordinates $\{\zeta\}$, $P_I(\zeta)$ are polynomials
(perhaps, of several variables), thus $dQ(\zeta) \sim
{\cal Q}'(\zeta) = \sum_K q_KP_K(\zeta)$. Also
$d\omega(\zeta) \sim {\cal P}'(\zeta)$,
$d\lambda(\zeta) \sim \dot{\cal P}(\zeta)$
with some polynomials
${\cal P}'(\zeta)$, $\dot{\cal P}(\zeta)$ and (\ref{WDVV})
is equivalent to a certain associative algebra of polynomials
\be
P_I P_J = {C^{(Q)}}_{IJ}^K P_K{\cal Q}'\ {\rm mod}\
({\cal P}',\dot{\cal P}).
\label{algpol}
\ee
In order to obtain the WDVV equations one still needs to express
the structure constants matrices $C_I^{(Q)}$  through $F_I$
(which, in contrast to $C_I$, are independent of the ``metric''
$Q$).  In all relevant cases the third derivatives of the
prepotential are expressed in terms of the same 1-differentials
as in (\ref{algdiff}):
\be
F_{IJK} = \stackreb{d\omega = 0}{\res}
\frac{dW_IdW_JdW_K}{d\lambda d\omega},
\label{2.10}
\ee
where $\delta\lambda \wedge \delta\omega$ is the symplectic form
on the phase space of $1d$ integrable system, associated with given
$4d$ model.

From (\ref{metric}) and (\ref{2.10}),
\be
Q_{LK} \equiv  \sum_M q_M F_{LKM} =
\stackreb{d\omega = 0}{\res} \frac{dW_LdW_KdQ}{d\lambda d\omega}
\ee
and according to (\ref{2.8})
\be
F_{IJK} = {C^{(Q)}}_{IJ}^LQ_{LK}.
\ee
Given (\ref{2.8}) and (\ref{2.10}) this provides the proof of the WDVV eqs.

We derive eq.(\ref{2.10}) in detail in sect.\ref{npsw} and consider
particular examples of the SW theory and associated algebras
(\ref{2.8}) in sect.\ref{Examples}. Let us now make
several comments.

Conventional derivations of the WDVV equations rely upon the
assumption that at least one of the ``metrics'' $Q$
in (\ref{WDVV}) is flat. Under this assumption one can relate
the associativity equation (\ref{WDVV}) to the flatness of
the connection
\be
\partial_I \delta^A_B + zC^A_{IB}
\ee
with {\it arbitrary} spectral parameter $z$, and then to the theory
of deformations of Hodge structures and the standard theory
of quantum cohomologies.
There is no {\it a priori} reason for any
\be
Q_{IJ} =  \stackreb{d\omega = 0}{\res}
\frac{dW_IdW_JdQ}{d\lambda d\omega},
\ee
to be flat and in this context such kind of derivation -- even if
exists -- would look somewhat artificial.

In our presentation the WDVV equations depend on a
{\it triple} of differentials on the spectral curve,
$d\omega$, $d\lambda$, $dQ$. The first two, $d\omega$ and $d\lambda$,
describe the symplectic structure $\delta\omega \wedge \delta\lambda$
of the relevant integrable system,
this is the pair, discussed in \cite{DKN,KriW,KriPho,M} and in another
context in \cite{KM}.  The choice of the third differential $dQ$
is more or less arbitrary.\footnote{
In the conventional Landau-Ginzburg topological models the spectral
curve is Riemann sphere. In this case $dQ$ can be considered as
``dressed'' $d\lambda$: $dQ(\lambda) = q(\lambda)d\lambda$
with some {\it polynomial} $q(\lambda)$ and there is a distinguished choice,
$q(\lambda)=1$, i.e. $dQ = d\lambda$ when the associated metric is
flat \cite{Losev,KriW}.
}
Sometime (for example, for hyperelliptic
spectral surfaces) the algebra (\ref{algdiff})
depends only on the pair $dQ$, $d\omega$ (or ${\cal Q}'$,
${\cal P}'$ in the formulation (\ref{algpol})).
We shall demonstrate below that just in this case the algebra (\ref{2.8}) is
associative and the WDVV equations are fulfilled.
The pair of differentials in this case is essentially the same
as arising in many related contexts: $(p,q)$
in minimal conformal models, a
pair of polynomials in associated matrix models \cite{Dou,FKN,KM}, a pair of
operators in the Kac-Schwarz problem \cite{Sch,KM}
and in a bispectral problem in the
theory of KP/Toda hierarchies \cite{OrHar} etc. The simplest example of this
structure
is the family of associative rings in the number theory, with multiplication
$ab = qc + pd$ with co-prime $q$ and $p>q$ where $a,b,c$ belong to the field
of residues modulo $p$, while $d$ is defined modulo $q$.  All the algebras
with multiplication rule $a\cdot_q b = c$ are isomorphic, but if considered
as a {\it family} over the ``moduli space'' of $q$'s, they  form some new
non-trivial structure.
In this oversimplified example the ``moduli space'' is discrete,
but it becomes continuous as soon as one switches to the families of the
polynomial rings (\ref{algpol}). One can assume that it is this structure -- a
{\it family} of associative algebras (not just a {\it single} associative
algebra) -- that stands behind the WDVV equations, and then it should be
searched for in the theory of quantum cohomologies, and in topological
theories.  (Let us emphasize \cite{MMM} that even if some particular metric
$Q$ is flat, this is normally not so for all other $Q$'s.)

\section{The SW prepotentials. Perturbative examples \label{pert}}
\setcounter{equation}{0}
\subsection{General formulas}
If the exact prepotential of the SW theory satisfies the WDVV
equations, the same should be true for its perturbative part
{\it per se}, since in $4d$ SUSY YM theory the perturbative contribution
to effective action is one-loop, pure logarithmic, and well
defined as a {\it leading} term in $1/a$ expansion.
In \cite{MMM} we discussed this phenomenon for the pure gauge ${\cal N}=2$
SUSY model and in this section we are going to present more general examples.

Perturbatively, the prepotential
of the pure gauge ${\cal N}=2$ SUSY YM model with the gauge group
$G=SU(N_c)$ is given by
\be
F_0 = \frac{1}{2}\sum_{1\leq r<r'\leq N_c}
(a_r-a_{r'})^2 \log(a_r-a_{r'}) =
\frac{1}{4}\sum_{r,r'}(a_r-a_{r'})^2 \log(a_r-a_{r'}),
\ \ \ \sum_{1\leq r \leq N_c} a_r = 0
\label{3.13}
\ee
This formula establishes that when v.e.v.'s
of the scalar fields in the gauge supermultiplet are non-vanishing
(perturbatively $a_r$ are eigenvalues of the vacuum
expectation matrix  $\langle\phi\rangle$), the fields in the gauge multiplet
acquire masses $m_{rr'} = a_r - a_{r'}$ (the pair of indices $(r,r')$ label
a field in the adjoint representation of $G$), and
\be\label{masses}
F =  \frac{1}{4}\sum_{masses}  (mass)^2 \log (mass).
\ee
Now, the formula (\ref{masses}) can be easily generalized.

There are two origins of masses in ${\cal N}=2$ SUSY YM models:
first, they can be generated by vacuum values of the scalar $\phi$ from  the
gauge supermultiplet. For a supermultiplet in representation $R$
of the gauge group $G$ this contribution to the prepotential
is given by the analog of the Coleman-Weinberg formula:
\be
F_R = \pm\frac{1}{4} \Tr_R \ \phi^2\log\phi,
\label{ColeWei}
\ee
and the sign is ``$+$'' for vector supermultiplets (normally they
are in the adjoint representation) and ``$-$'' for matter hypermultiplets.
Second, there are bare masses $m_R$ which should be added to $\phi$
in (\ref{ColeWei}). As a result, the general expression for the
perturbative prepotential is
\be
F = \frac{1}{4}\sum_{vector\ multiplets} \Tr_{A}
(\phi + M_nI_A)^2\log(\phi + M_nI_A) - \nn \\
- \frac{1}{4}\sum_{hypermultiplets} \Tr_R
(\phi + m_RI_R)^2\log(\phi + m_RI_R) + f(m)
\label{PertuF}
\ee
where the term $f(M)$ depending only on the masses is not fixed by the
(perturbative) field theory but can be read off from the
non-perturbative description and $I_R$ denotes the unit matrix in
representation $R$.

Now let us consider some examples. Sometimes, when explicit calculations are
tedious, we have used the computer simulations with the MAPLE program to
check the perturbative formulas for the groups of low
ranks. The proof of these formulas comes from the consideration of
sect.5.2. We start with the gauge group $SU(N_c)$ and
consider its different representations.

\subsection{$A_n$ models}

Formula (\ref{PertuF}) can be immediately rewritten in terms of
the eigenvalues $a_i$ of $\langle\phi\rangle _F$ in the fundamental
representation $F$. Since the adjoint representation
$A \in F\times\bar F$, the eigenvalues of the same
$\langle\phi\rangle$ in the adjoint representation are equal to
$a_r - a_{r'}$. Thus,
the contribution of the adjoint hypermultiplet is equal to
\be\label{FA}
F_{A} = \pm\frac{1}{4} \sum_{r,r'} (a_r-a_{r'}+M)^2
\log(a_r-a_{r'}+M),
\ee
in accordance with the original formula (\ref{3.13}).
The contribution of each fundamental hypermultiplet is
\be
F_{F} = -\frac{1}{4} \sum_r (a_r + m)^2\log(a_r+m).
\ee
For the hypermultiplet in the symmetric/antisymmetric square of the
fundamental representation
\be
F_{S} =  -\frac{1}{4}\sum_{r\leq r'} (a_r+a_{r'}+m)^2
\log(a_r+a_{r'} +m), \nn \\
F_{AS} =  -\frac{1}{4}\sum_{r < r'} (a_r+a_{r'}+m)^2
\log(a_r+a_{r'} +m).
\ee
In further consideration we change the sign of masses to make them more
similar to variables $a$'s.

Normally, in $4d$ model there is only one vector multiplet in the
adjoint representation of the gauge group.
In order to preserve the asymptotic freedom, the set of hypermultiplets
should satisfy
\be
N_c - \sum_R N_RT_R \geq 0,
\ee
where $n_R$ is the number of hypermultiplets in representation $R$
and $\Tr_R t_at_b = T_R\delta_{ab}$ (i.e. $T_F = 1/2$, $T_A = N_c$,
$T_S = \frac{N_c+2}{2}$, $T_{AS} = \frac{N_c-2}{2}$). Thus, there are
six choices for the UV-finite models:
\be
N_A=1, \nn \\
N_F=2N_c, \nn \\
N_S = N_{AS} = 1, \nn \\
N_S=1,\ \  N_F=N_c-2, \nn \\
N_{AS}=1,\ \  N_F = N_c+2, \nn \\
N_{AS}=2, \ \ N_F = 4,
\ee
and four series of models with dynamically generated scale
$\Lambda_{QCD}$,
\be
N_F = 0,1,\ldots,2N_c-1, \nn \\
N_S = 1,\ \  N_F = 0,1,\ldots, N_c-3, \nn \\
N_{AS} = 1,\ \  N_F = 0,1,\ldots, N_c+1, \nn \\
N_{AS} = 2, \ \ N_F = 0,1,2,3.
\ee

In \cite{MMM} we checked explicitly that (\ref{3.13})
satisfies the WDVV eqs. Simple computer simulations imply
that (\ref{PertuF}) is also consistent with the WDVV eqs,
at least, if there is only one vector multiplet, if
all hypermultiplets are in the (first) fundamental representation
of the gauge group $SU(N_c)$ and if (cf. with (\ref{masses}))
\be
f_F(m) = {1\over 4} \sum _{\iota,\iota'}(m_{\iota} - m_{\iota'})^2
\log (m_\iota - m_{\iota'})
\ee
Generally, if one considers the prepotential of the form
\be\label{26}
F_0+rF_F+Kf_F(m)
\ee
the WDVV equations are satisfied with arbitrary $r$ and $K=r^2/4$. Note
that for $r=-2$ one can consider (\ref{26}) as a perturbative expression with
the "extended" gauge group $SU(N_c+N_F)$.

As for the other possible representations of $SU(N_c)$, the prepotential
satisfies the WDVV equations iff $r$ is adjusted so that the leading term at
large $a$ is zero. This is equivalent to the condition of the ultraviolet
finiteness (or zero $\beta$-function), but without special matching of the
multiplet number the coefficient $r$ is non-integer. Say, for $N_S$
hypermultiplets in the symmetric representation
$r=T_A/N_ST_S={2N_c\over N_S(N_c+2)}$.

The WDVV equations, however, are always fulfilled for the symmetric and
antisymmetric representations with {\it zero} masses. We will discuss a little
later a general empirical rule that allows one to determine whether the
prepotential satisfies the WDVV eqs.

\subsection{$B_n$, $C_n$, $D_n$ models}

Now let us turn to the other groups. The eigenvalues of $\langle\phi\rangle$
in the first
fundamental representation of the classical series of the Lie groups are
\be
B_n\ (SO(2n+1)):\ \ \ \ \ \{a_1,...,a_n,0,-a_1,...,-a_n\};\\
C_n\ (Sp(n)):\ \ \ \ \ \{a_1,...,a_n,-a_1,...,-a_n\};\\
D_n\ (SO(2n)):\ \ \ \ \ \{a_1,...,a_n,-a_1,...,-a_n\}
\ee
while the eigenvalues in the adjoint representation have the form
\be\label{adj}
B_n:\ \ \ \ \ \{\pm a_j;\pm a_j\pm a_k\};\ \ \ j<k\le n\\
C_n:\ \ \ \ \ \{\pm 2a_j;\pm a_j\pm a_k\};\ \ \ j<k\le n\\
D_n:\ \ \ \ \ \{\pm a_j\pm a_k\}, \ \ j<k\le n
\ee
Analogous formulas can be written for the exceptional groups too. Then,
the WDVV equations in the perturbative regime should be true for the pure
gauge theory with any group (including exceptional ones)
\footnote{The rank of
the group should be bigger than 2 for the WDVV equations not to be empty,
thus for example in the pure gauge $G_2$-model they are satisfied trivially.
We have checked that the WDVV
equations hold for the perturbative prepotential of the pure gauge $F_4$
model. Note that the corresponding non-perturbative SW curve is not
obviously hyperelliptic. Thus, our general reasoning in sect.4 is not directly
applicable to this case.}. The prepotential in the pure gauge theory can be
read off from the formula (\ref{adj}) and has the form
\be\label{adjvo}
B_n:\ \ \ \ \ F_0={1\over 4}\sum_{i,i} \left(\left(a_{i}-a_{j}
\right)^2\log\left(a_{i}-a_{j}\right)+\left(a_{i}+a_{j}
\right)^2\log\left(a_{i}+a_{j}\right)\right)+
{1\over 2}\sum_i a_i^2\log a_i;\\
C_n:\ \ \ \ \ F_0={1\over 4}\sum_{i,i} \left(\left(a_{i}-a_{j}
\right)^2\log\left(a_{i}-a_{j}\right)+\left(a_{i}+a_{j}
\right)^2\log\left(a_{i}+a_{j}\right)\right)
+2\sum_i a_i^2\log a_i;\\
D_n:\ \ \ \ \ F_0={1\over 4}\sum_{i,i} \left(\left(a_{i}-a_{j}
\right)^2\log\left(a_{i}-a_{j}\right)+\left(a_{i}+a_{j}
\right)^2\log\left(a_{i}+a_{j}\right)\right)
\ee
Let us prove that these prepotentials do really satisfy the WDVV equations.
The proof is much similar to that presented in paper \cite{MMM}. We start
with the general prepotential of the form
\be
F={1\over 4}\sum_{i,i} \left(\alpha_-\left(a_{i}-a_{j}
\right)^2\log\left(a_{i}-a_{j}\right)+\alpha_+\left(a_{i}+a_{j}
\right)^2\log\left(a_{i}+a_{j}\right)\right)+{\eta\over 2}\sum_i a_i^2\log a_i
\ee
$\alpha_{\pm}$ and $\eta$ being arbitrary constants. Then,
\be\label{f}
\{(F_i)_{mn}\} = \frac{\delta_{mn}(1-\delta_{mi})(1-\delta_{ni})}
{\alpha_{im}} - \frac{\delta_{mi}(1-\delta_{ni})}{\alpha_{in}}
- \frac{\delta_{ni}(1-\delta_{mi})}{\alpha_{im}} + \nn \\ +
\left(\frac{1}{\alpha_i} + \sum_{l\neq i}\frac{1}{\alpha_{ik}}\right)
\delta_{mi}\delta_{ni}
\ee
where
\be
\alpha_{mk}\equiv {a_m^2-a_k^2\over \alpha_+(a_m-a_k)+\alpha_-(a_m+a_k)},
\ \ \ \ \alpha_k\equiv {a_k\over\eta}
\ee
The inverse matrix
\be\label{f-1}
\{(F_k^{-1})_{mn}\} = \alpha_k + \delta_{mn}\alpha_{km}(1-\delta_{mk}),
\ee
Now in order to prove the WDVV equation, it is sufficient to check the
symmetricity w.r.t. the indices of the expression $(F_i^{-1}F_jF_k^{-1})_{
\alpha\beta}$. From formulas (\ref{f})-(\ref{f-1}), one can
obtain that
\be\label{long}
(F_i^{-1}F_jF_k^{-1})_{\alpha\beta}={\alpha_i\alpha_k\over \alpha_j}+
\delta_{\alpha\beta}(1-\delta_{i\alpha})(1-\delta_{k\alpha})
(1-\delta_{j\alpha}){\alpha_{i\alpha}\alpha_{k\beta}\over \alpha_{j\beta}}+
\delta_{j\alpha}\delta_{j\beta}(1-\delta_{i\alpha})(1-\delta_{k\beta})
\left({1\over \alpha_j}+\sum_{n\ne j}{1\over \alpha_{jn}}\right)+\\
+\delta_{j\alpha}(1-\delta_{i\alpha})\alpha_{i\alpha}\left(
{\alpha_k\over \alpha_j}-{\alpha_{k\beta}\over
\alpha_{j\beta}}(1-\delta_{k\beta})
(1-\delta_{j\beta})\right)+\delta_{j\beta}(1-\delta_{k\beta})
\alpha_{k\beta}\left(
{\alpha_i\over \alpha_j}-
{\alpha_{i\alpha}\over \alpha_{j\alpha}}(1-\delta_{i\alpha})
(1-\delta_{j\alpha})\right)=\\
={\alpha_i\alpha_k\over \alpha_j}+
\delta_{\alpha\beta}(1-\delta_{i\alpha}-\delta_{k\alpha}
-\delta_{j\alpha}){\alpha_{i\alpha}\alpha_{k\beta}\over \alpha_{j\beta}}+
\delta_{j\alpha}\delta_{j\beta}\left({1\over \alpha_j}+
\sum_{n\ne j}{1\over \alpha_{jn}}\right)+\\
+\delta_{j\alpha}\alpha_{i\alpha}\left(
{\alpha_k\over \alpha_j}-{\alpha_{k\beta}\over
\alpha_{j\beta}}(1-\delta_{k\beta}
-\delta_{j\beta})\right)+\delta_{j\beta}\alpha_{k\beta}\left(
{\alpha_i\over \alpha_j}-
{\alpha_{i\alpha}\over \alpha_{j\alpha}}(1-\delta_{i\alpha}
-\delta_{j\alpha})\right)
\ee
where it is used that $i\ne j\ne k$. The first three terms are evidently
symmetric with respect to interchanging $\alpha\leftrightarrow\beta$. The
 symmetricity of the last two terms implies the identity
\be
\alpha_{ij}\left({\alpha_k\over\alpha_j}-{\alpha_{k\beta}\over
\alpha_{j\beta}}\right)=
\alpha_{kj}\left({\alpha_i\over\alpha_j}-{\alpha_{i\beta}\over
\alpha_{j\beta}}\right)
\ee
This relation is identically satisfied with any $\eta$, but {\it only} at
$\alpha_+=\alpha_-$ or $\alpha_+=0$. These values exactly correspond
to the prepotentials for the classical groups (\ref{adjvo}) .

Consider now the theory with the gauge group $B_n$, $C_n$ or $D_n$
and the matter hypermultiplets in the first fundamental representation. Then,
one again needs to add to the prepotential some appropriate
terms $f_F(M)$ which
depend only on masses and can not be determined from perturbative quantum field
theory. The correct form of these terms can be obtained from the requirement
for the prepotential to satisfy the WDVV equations; the result reads
\be\label{adjot}
F=F_0+rF_F+{r^2\over 16} \sum_{\iota,\iota'}\left(\left(m_{\iota}-m_{\iota'}
\right)^2\log\left(m_{\iota}-m_{\iota'}\right)+\left(m_{\iota}+m_{\iota'}
\right)^2\log\left(m_{\iota}+m_{\iota'}\right)\right)+
\nn \\
+ {r(r+s)\over 4}
\sum_{\iota}m_{\iota}^2\log m_{\iota}
\ee
where $s=2$ for the $B_n$ and $D_n$ series and to $s=-2$ for the $C_n$
series. This formula can be also obtained from the perturbative limit of the
non-perturbative theory, see sect.5.2. Note that, again at the value
$r=-2$, the prepotential (\ref{adjot}) is associated with the group of higher
rank. At the same time, at $r=2$ and $G=C_n$ one can consider $f(m)$ as
coming from the vector multiplet of the group $D_{N_f}$ and, inversely,
at $r=2$ and $G=D_n$ $f(m)$ comes from $C_{N_f}$.

The last $4d$ example, which is also of great importance, is the theory with
matter in the adjoint representation. This theory is associated with
the Calogero model (which is analyzed non-perturbatively in detail in
sect.6). In this case, the mass $M$ of the hypermultiplet
is not included into the set of moduli (otherwise the conformal covariance
would make the matrices in the WDVV eqs degenerate and non-invertible).
Keeping this in mind, one can easily check that the corresponding expression
(\ref{PertuF}) does {\it not} satisfy WDVV eqs. This result can be also
confirmed non-perturbatively, and we reveal its origin in the detailed
consideration of sect.6.

To formulate an empiric rule, one may say that the WDVV
equations are satisfied by perturbative prepotentials which depend only on
the sums of the type $(a_i\pm b_j)$, where moduli $a_i$ and $b_j$ are
either periods or masses. This is the case for the models that contain either
massive matter hypermultiplets in
the first fundamental representations (or their dual), or massless
matter in the square product of those.
Troubles arise in all other situations because of the terms
with $a_i\pm b_j\pm c_k\pm\ldots$\footnote{The inverse statement is wrong --
there are some exceptions when the WDVV equations hold despite the presence
of such terms -- e.g., for the exceptional groups.}

This general rule can be understood in terms
of the residue formulas, since it reflects the typical structures arising on
the sphere with punctures, which corresponds to the perturbative limit of
the non-perturbative hyperelliptic
curve as explained in sect.5.2. It is also consistent
with the D-brane description of the SW theory and implies that the WDVV
equations
may have some underlying structure behind this. This rule can
be also easily interpreted in D-brane terms, since the interaction of branes
is caused by strings between them. The pairwise structure $(a_i\pm
b_j)$ exactly reflects this fact, $a_i$ and $b_j$ should be identified with
the ends of string.

Let us note that the non-pairwise terms in the prepotential drastically
destroy the specific structure of matrices (\ref{f})-(\ref{f-1}),
spoiling the WDVV equations.

\subsection{Beyond $4d$}

This connection can be extended even further if one enlarges
the set of multiplets and adds more vector multiplets
(which all {\it enhance} the asymptotic freedom). A natural way to
do this is to consider some compactification of YM theory from higher
dimensions: then the Kaluza-Klein mechanism will produce an infinite
tower of vector multiplets in the adjoint representation, with
masses being all dependent on the finite number of parameters
of the compact manifold. In the simplest case of toroidal compactifications,
one has:
\be
\{M\} = Mn, \ \ n = 0,1,2,\ldots
\ee
for compactification from $5d$ onto a circle \cite{N}, or
\be
\{M\} = M_1n_1 + M_2n_2, \ \ n_1,n_2 = 0,1,2,\ldots
\ee
for compactification from $6d$ onto a torus.
In these cases the formula (\ref{PertuF}) would contain infinite sums
and require some regularization.
As a result, in the $5d$ case, the second derivative of the
prepotential behaves like $\log(\sinh{a\over M})$, i.e. the prepotential
is of the form $M^2Li_3\left({e^{a}\over M}\right)$, $Li_3\left(e^{x}
\right)\equiv\sum_{k=1} {e^{kx}\over k^3}$, $Li_3''\left(e^x\right)=
-\log\left(1-e^{x}\right)=-{x\over 2}-\log (2\sinh x/2)$, while,
in the $6d$ case, the second derivative is rather logarithm of the
Jacobi theta-function $\left.\log\left[\theta\left({a\over M_2}| {M_1\over
M_2}\right)\right]\right.$ and the corresponding prepotential can be
represented as a sum of three-logarithms. String theory associates new
degrees of freedom with wrapping of string (and brane) configurations around
the compactified directions and more sophisticated sums of three-logarithms
appear in the expression for the prepotential.

Let us look at the $5d$ theory more carefully. From now on, we normalize
$a_i$ so that the mass parameter $M$ is equal to 1.
One can easily check that the
WDVV equations are {\em not} satisfied by naive three-logarithm prepotential
discussed in the literature \cite{N}.
However, this situation can be
corrected by adding some cubic terms unexpected from the field theory
point of view. These terms are equal to $\sum_{i>j>k}a_ia_j
a_k$ so that the complete perturbative prepotential satisfying the WDVV eqs
is
\be\label{FARTC}
F={1\over 4}\sum_{i,j}\left({1
\over 3}a_{ij}^3-{1\over 2}Li_3\left(e^{-2a_{ij}}\right)\right)-
{N_c\over 4}\sum_{i>j>k}a_ia_ja_k
\ee
where $a_{ij}\equiv a_i-a_j$.
The cubic terms in (\ref{FARTC}) could be compared with the $U^3$-terms in
the perturbative part of
the prepotential of heterotic string, coming from the
requirement of modular invariance \cite{HM}. We show in sect.5.2 how these
terms can be obtained in the
perturbative limit of the SW non-perturbative description of $5d$ theory.

\section{Non-perturbative proof of WDVV equations \label{npsw}}
\setcounter{equation}{0}
\subsection{SW theory and the general theory of Riemann surfaces}
Non-perturbatively, the SW construction implies that
for any $4d$ ${\cal N}=2$ SUSY Yang-Mills model one associates
a (finite-dimensional) integrable system, its phase space is a family of
abelian varieties (Hamiltonian tori) over a moduli space ${\cal M}$,
parametrized by the angle and action variables respectively.
The phase space itself is directly "observable" when the $4d$
model is compactified to $3d$, i.e. defined in the space-time
$R^3\times S^1$ rather than $R^4$: then the angle variables
of the integrable system are identified with the Wilson loops along
the compactified dimension in the $4d$ model \cite{SW3}.
In the non-compactified case the only four-dimensional observables
are the periods of the generating 1-form $dS="pdq"$
along the non-contractable contours on the Hamiltonian tori. The period
matrix for the given system consists of the second derivatives of the
prepotential while the $B$-periods of $dS$ are its first derivatives.
$A$-periods of $dS$ define the {\it flat} structure on moduli space.

In the context of SW theory one actually deals with specific
integrable systems associated with Riemann surfaces
(complex curves). This can be understood from the point of view of
string theory, which relates the emergency of Yang-Mills theories
from string compactifications to emergency of Hitchin-like integrable
systems from more general systems \cite{DoMa} associated with the Calabi-Yau
families.  In practice it means that the families of abelian varieties,
relevant for the SW theory are in fact the Jacobian tori of complex curves.

In the context of SW theory one starts with a {\it bare} spectral curve,
with a holomorphic 1-form $d\omega $, which is elliptic curve (torus)
\be\label{torus}
E(\tau):\ \ \ \ y^2 = \prod_{a=1}^3 (x - e_a(\tau)), \ \ \
\sum_{a=1}^3 e_a(\tau) = 0,\ \ \ \ \ d\omega = \frac{dx}{y},
\ee
when the YM theory is UV finite, or its degeneration $\tau
\rightarrow i\infty$ -- the double-punctured sphere (``annulus''):
\be
x\rightarrow w\pm\frac{1}{w},\ \
y \rightarrow w\mp\frac{1}{w},\ \ \ \ \ d\omega = \frac{dw}{w}
\ee
From the point of view of integrable system, the Lax operator
${\cal L}(x,y)$ is defined as a function (1-differential) on the
{\it bare} spectral curve \cite{KriCal}, while the {\it full} spectral curve
${\cal C}$ is given by the Lax-eigenvalue equation: $\det({\cal L}(x,y) -
\lambda ) = 0$. As a result, ${\cal C}$ arises as a ramified covering over the
{\it bare} spectral curve:
\be
{\cal C}:\ \ \ {\cal P}(\lambda; x,y) = 0
\ee
In the case of the gauge group  $G=SU(N_c)$, the function ${\cal P}$ is a
polynomial of degree $N_c$ in $\lambda$.

The function ${\cal P}$ depends also on parameters (moduli)
$s_I$, parametrizing the moduli space ${\cal M}$. From the
point of view of integrable system, the Hamiltonians
(integrals of motion) are some specific co-ordinates on
the moduli space. From
the four-dimensional point of view, the co-ordinates  $s_I$ include $s_i$ --
(the Schur polynomials of) the adjoint-scalar expectation values $h_k =
\frac{1}{k}\langle\Tr \phi^k\rangle$ of the vector ${\cal N}=2$
supermultiplet, as well as $s_\iota = m_\iota$ -- the masses of the
hypermultiplets.

The generating 1-form $dS \cong \lambda d\omega$ is meromorphic on
${\cal C}$ (hereafter the equality modulo total derivatives
is denoted by ``$\cong$''). The prepotential is defined in terms of the
cohomological class of $dS$:
\be
a_I = \oint_{A_I} dS, \ \ \ \ \ \
\frac{\partial F}{\partial a_I} = \int_{B_I} dS \nn \\
A_I \circ B_J = \delta_{IJ}.
\label{defprep}
\ee
The cycles $A_I$ include the $A_i$'s wrapping around the handles
of ${\cal C}$ and $A_\iota$'s, going around the singularities
of $dS$ (for simplicity we assume that they are
simple poles, generalization is obvious).
The conjugate contours $B_I$ include the cycles $B_i$ and the
{\it non-closed} contours $B_\iota$, ending at the singularities
of $dS$ (see sect.5 of \cite{IM2} for more details).
The integrals $\int_{B_\iota} dS$ are actually divergent, but
the coefficient of divergent part is equal to residue of $dS$
at particular singularity, i.e. to $a_\iota$. Thus the divergent
contribution to the prepotential is quadratic in $a_\iota$, while
the prepotential is normally defined {\it modulo} quadratic combination
of its arguments (since only its {\it third} derivatives appear in the
WDVV equations). In
particular models $\oint_{A,B} dS$ for some conjugate pairs of contours are
identically zero on entire ${\cal M}$: such pairs are not included into our
set of indices $\{I\}$.

In the context of SW theory with the data $\{ {\cal M}, {\cal C}, dS\}$
the number of independent $a_I$ is exactly the same as ${\rm dim}_C{\cal M}$.
This allows us to use the functions $a_I(s)$ as co-\-ordinates on
moduli space ${\cal M}$.

The self-consistency of the definition (\ref{defprep}) of $F$,
i.e. the fact that the matrix
$\frac{\partial^2 F}{\partial a_I\partial a_J}$ is symmetric is guaranteed
by the following reasoning.
Let us take the derivative of $dS$ over moduli:
\be
\frac{\partial dS}{\partial s_I} \cong \frac{\partial\lambda}
{\partial s_I} d\omega = -\frac{\partial {\cal P}}{\partial s_I}
\frac{d\omega}{{\cal P}'} \equiv dv_I.
\label{defdv}
\ee
where prime denotes the $\lambda$-derivative (for constant $x, d\omega$ and
$s_I$).
Note that we did not include $\tau$ into the set of moduli
$\{s_I\}$, therefore we can take $s_I$-derivatives keeping $(x,y)$
(and thus $d\omega$) fixed: the variation of coordinate on bare curve would
not change the cohomology classes.
The variation of ${\cal P}(\lambda; x,y) = 0$
\be
{\cal P}'d\lambda + \frac{\partial{\cal P}}{\partial \omega}
d\omega = 0,
\ee
implies that the zeroes of $d\omega$ and ${\cal P}'$ on ${\cal C}$
coincide (since $d\lambda$ and $d\omega$ are supposed to have non-coinciding
zeroes), and, therefore, $\frac{d\omega}{{\cal P}'}$ is a {\it holomorphic}
1-differential on ${\cal C}$. If $\partial{\cal P}/\partial s_I$ is
non-singular, $dv_I$ is also holomorphic\footnote{Since curves with punctures
and the corresponding meromorphic differentials can be obtained by
degeneration of smooth curves of higher genera we do not make any distinction
between punctured and smooth curves below. We remind that the holomorphic
1-differentials can have at most simple poles at the punctures while
quadratic differentials can have certain double poles etc.}
 -- this is the case of $dv_i$, while
$dv_\iota$ have poles encircled by contours $A_\iota$.  The second derivative
\be
\frac{\partial^2 F}{\partial a_I\partial a_J} = F_{IJ}
\ee
is essentially the period matrix of the punctured Riemann surface
${\cal C}$
\be
\sum_J F_{IJ}\oint_{A_J} dv_L=\int_{B_I} dv_L +\int_{{\d B_I\over \d s_L}}
dS
\ee
The second term at the r.h.s. can be non-vanishing if the contour $B_I$ is
not closed. The end-points of such contours (punctures) are included into the
set of moduli (this actually implies that a local coordinate is specified in
the vicinity of the puncture). The derivative ${\d B_I\over \d s_L}$ is
actually a point, thus, $\int_{{\d B_I\over \d s_L}} dS={dS\over
d\lambda}\left(\lambda= {\d B_I\over \d s_L}\right)$. This term also cancels
the possible divergency in $\int_{B_I} dv_L$, making $F_{IJ}$ finite.

As any period matrix $F_{IJ}$ is symmetric:
\be
\sum_{IJ} (F_{IJ} - F_{JI}) \oint_{A_I}dv_K\oint_{A_J}dv_L =\\=
\sum_I\left(\oint_{A_I}dv_K\int_{B_I}dv_L -
\int_{B_I}dv_K\oint_{A_I}dv_L\right)
+ \sum_I\left(\oint_{A_I}dv_K \int_{{\d B_I\over \d s_L}}
- \int_{{\d B_I\over \d s_K}} \oint_{A_I}dv_L\right)
\ee
The first sum at the r.h.s. is equal to
\be
\sum_I\left(\oint_{A_I}dv_K\int_{B_I}dv_L -
\int_{B_I}dv_K\oint_{A_I}dv_L\right)=
{\res} \left( v_K dv_L \right) = 0
\ee
since all the singularities of $dv_K$ are already taken into account
by inclusion of $A_\iota$ into our set of $A$-cycles. The second sum is also
zero because the only non-vanishing entries are those with $K=L$.

In this paper we also need the moduli derivatives of the prepotential.
It is easy to get:
\be
\sum_{IJ} \frac{\partial F_{IJ}}{\partial s_M}
\oint_{A_I}dv_K\oint_{A_J}dv_L =
\sum_I\left(\oint_{A_I}dv_K\int_{B_I}\frac{\partial dv_L}{\partial s_M} -
\int_{B_I}dv_K\oint_{A_I}\frac{\partial dv_L}{\partial s_M}\right) +\Sigma
_{KLM}=
\nn \\ =
{\res} \left(dv_K \frac{\partial v_L}{\partial s_M}\right)+\Sigma_{KLM}
\label{prom1}
\ee
where
\be\label{sigma}
\Sigma_{KLM}=\sum_I\left(\oint_{A_I}dv_K {\p s_M}\int_{{\d B_I\over \d
s_L}}dS+ \oint_{A_I}dv_K \int_{{\d B_I\over \d s_M}}dv_L-
\oint_{A_I}{\partial dv_L\over\partial s_M} \int_{{\d B_I\over \d s_K}}dS
\right)
\ee
is the contribution of the moduli dependence of punctures.

The residue in (\ref{prom1}) does not need to
vanish, since the derivatives w.r.t. the
moduli produces new singularities. Indeed, from (\ref{defdv}) one obtains
\be
-\frac{\partial dv_L}{\partial s_M} =
\frac{\partial^2{\cal P}}{\partial s_L\partial s_M}
\frac{d\omega}{{\cal P}'} +
\left(\frac{\partial{\cal P}}{\partial s_L}\right)'
\left(-\frac{\partial{\cal P}}{\partial s_M}\right)
\frac{d\omega}{{\cal P}'} -
\frac{\partial{\cal P}}{\partial s_L}
\frac{\partial{\cal P}'}{\partial s_M}
\frac{d\omega}{({\cal P}')^2}  +  \frac{\partial{\cal P}}{\partial s_L}
\frac{\partial{\cal P}}{\partial s_M}
\frac{{\cal P}''d\omega}{({\cal P}')^3}=
\nn \\
= \left[
\left(\frac{ {\partial{\cal P}/\partial s_L}
 {\partial{\cal P}/\partial s_M} }{ {\cal P}'}  \right)' +
\frac{\partial^2{\cal P}}{\partial s_L\partial s_M} \right]
\frac{d\omega}{{\cal P}'},
\ee
and this expression has new singularities (second order poles)
at the zeroes of ${\cal P}'$ (i.e. of $d\omega$). Note,
that the contributions from the singularities of
$\partial{\cal P}/\partial s_L$, if any, are already taken into
account in the l.h.s. of (\ref{prom1}).
Picking up the coefficient at the leading singularity, we obtain:
\be\label{resch}
{\res} \left(dv_K \frac{\partial v_L}{\partial s_M}\right) +\Sigma_{KLM}=
-\stackreb{d\omega = 0}{\res}
\frac{\partial{\cal P}}{\partial s_K}
\frac{\partial{\cal P}}{\partial s_L}
\frac{\partial{\cal P}}{\partial s_M}
\frac{d\omega^2}{({\cal P}')^3 d\lambda} =
\stackreb{d\omega = 0}{\res}
\frac{dv_Kdv_Ldv_M}{d\omega d\lambda}
\ee
The integrals at the l.h.s. of (\ref{prom1}) serve to convert
the differentials $dv_I$ into {\it canonical} ones $dW_I$,
satisfying $\oint_{A_I} dW_J = \delta_{IJ}$. The same matrix
$\oint_{A_I} dv_J$ relates the derivative w.r.t. the moduli $s_I$
and the periods $a_I$. Putting all together we obtain
(see also \cite{KriW}):
\be
\frac{\partial F_{IJ}}{\partial s^K} =
\stackreb{d\omega = 0}{\res}
\frac{dW_IdW_Jdv_K}{d\omega d\lambda}; \nn \\
\frac{\partial^3 F}{\partial a^I\partial a^J\partial a^K} =
\frac{\partial F_{IJ}}{\partial a^K} =
\stackreb{d\omega = 0}{\res}
\frac{dW_IdW_JdW_K}{d\omega d\lambda}
\label{resfor}
\ee

\subsection{Algebra of 1-differentials \label{algebdiff}}

We consider particular examples of the algebra (\ref{2.8}) in
sect.\ref{Examples}. Here we sketch some common features
of these examples that seem to be relevant for existence of such
algebras and show that they always exist in the hyperelliptic case.

{\bf 1)} The linear space of differentials $dW_I$ in (\ref{2.8})
(or the space of $dv_I$ in (\ref{defdv})) differs a little bit from the set
of {\it all} holomorphic 1-differentials\footnote{For the gauge
groups $B_n$, $C_n$, $D_n$ one should take {\it all} 1-differentials which
are {\it odd} under the involution $\lambda\to -\lambda$.}
on punctured
\footnote{We mean here by
"punctured" that the "holomorphic differentials" can have at most {\it
simple} poles at punctures.  } spectral curve ${\cal C}$ for the following
reasons. In elliptic (UV-finite) cases, when the bare spectral curve is
$E(\tau)$, $d\omega$ is holomorphic on ${\cal C}$, but it is not included
into the set $\{dW_I\}$, because the algebra of $dW_I$ is defined {\it
modulo} $d\omega$.  Up to such possible corrections, the number of linearly
independent $dW_I$'s is $g + n$, where $g$ is the genus of ${\cal C}$ and
$n+1$ -- the number of punctures.

{\bf 2)}
The products $dW_IdW_J$ are holomorphic quadratic differentials
on the punctured curve ${\cal C}$. There are $\frac{(g+n)(g+n-1)}{2}$
(or $\frac{(g+n-1)(g+n-2)}{2}$) such products, but at most
$3g-3+2n$ of them are linearly independent, since this is the
quantity of independent quadratic differentials.

{\bf 3)} We want to obtain an associative algebra of 1-differentials
$dW_I$,
\be
dW_I\circ dW_J = \sum_K C_{IJ}^K dW_K
\ee
which is an ordinary algebra of multiplications
{\it modulo} $d\omega$ and $d\lambda$.
In addition to these two 1-differentials one should also
specify $dQ \equiv \sum_I q_IdW_I$.
Then, any quadratic differential can be represented as their ``linear
combination'', namely:
\be
dW_I dW_J = \left(\sum_K C_{IJ}^K dW_K\right) dQ
+ d\widetilde W_{IJ}^\omega d\omega +
d\widetilde W_{IJ}^\lambda d\lambda
\label{algmult}
\ee
and existence of such a representation follows already from
the estimation of the number of independent adjustment parameters.

For example, if there are no punctures, all $dW_I$ are holomorphic (have no
poles) and so does $d\omega$ (which is not included into the set $\{dW_I\}$),
while $d\lambda$ has the second order pole in a single point (this is
literally the case for the Calogero model, see sect.6), then $d\widetilde
W_{IJ}^Q \equiv \sum_K C_{IJ}^K dW_K$ has $g-1$ adjustment parameters ($g-1$
structure constants $C_{IJ}^K$ with $K = 1,2,\ldots,g-1$), while $d\widetilde
W_{IJ}^\omega$ and $d\widetilde W_{IJ}^\lambda$ are holomorphic
differentials, the latter one has the second order zero at the puncture,
i.e. they form $g$- and $g-2$-parametric families respectively. Thus,
totally we get $(g-1) +g+ (g-2)$ adjustment parameters, exactly exhausting
the $3g-3$-dimensional set of independent holomorphic quadratic $dW_IdW_J$.
Accurate analysis shows that the matching is preserved at punctured surface
as well.

\subsection{Associativity and the WDVV equations}
However, the existence of the closed algebra does not immediately imply its
associativity. Indeed, to check associativity one needs to consider the
expansion of products of the 3-differentials:
\be\label{47}
dW_IdW_JdW_K=dW_{IJK}dQ^2+d^2\widetilde{\widetilde W^{\omega}
\!\!\!}_{_{IJK}}d\omega+
d^2\widetilde{\widetilde W^{\lambda}\!\!\!}_{_{IJK}}d\lambda
\ee
where $dW_{IJK}dQ$ is a set of holomorphic 1-differentials not
containing $d\omega$. Now repeating the above counting, we get that there
exist totally $(g-1)+(2g-1)+(3g-5)$, since $d^2\widetilde{\widetilde
W^{\omega}\!\!\!
}_{_{IJK}}$ are holomorphic quadratic differentials that do not contain
$d\lambda$, and $d^2\widetilde{\widetilde W^{\lambda}\!\!\!
}_{_{IJK}}$ are holomorphic
quadratic differentials with second order zeroes. Thus, for the $5g-5$
independent holomorphic 3-differentials we have too many, namely, $6g-7$
parameters.  Therefore, among the differentials $dW_{IJK}dQ$, $d^2\widetilde
{\widetilde W^{\lambda}\!\!\!}_{_{IJK}}$ and $d^2\widetilde
{\widetilde W^{\lambda}\!\!\!}_{_{IJK}}$ there are some relations that
generally spoil associativity. We present in sect.6 the explicit example
where associativity is indeed violated.

The situation simplifies different in the hyperelliptic case when there is
the additional ${\bf Z}_2$-transformation $\sigma: y\to -y$. In this case,
the holomorphic differentials, ${P_k(\lambda)d\lambda\over Y}$,
$k=0,\ldots g-1$ ($P_k$ denotes a polynomial of degree $k$) are {\it odd}
under $\sigma$ and their quadratic combinations, ${P_k(\lambda)
(d\lambda)^2\over Y^2}$,
$k=0,\ldots,2g-2$ are {\it even}.
The remaining $g-2$ quadratic holomorphic differentials, ${P_k(\lambda)
(d\lambda)^2\over
Y}$, $k=0,\ldots,g-3$ are $\sigma$-odd and do not appear in the l.h.s. of
(\ref{algmult}). On the other hand, in the expansion (\ref{algmult}),
one can throw away the term with $d\lambda$ at the r.h.s., since $d\lambda$
is $\sigma$-even and, therefore, the coefficient in front of it
is not a linear combination of the holomorphic differentials $dW_I$. This
restores the correct counting of $2g-1$ adjusting parameters for $2g-1$
differentials. It turns out that this restricted algebra spanned by the two
differentials $dQ$ and $d\omega$ is both closed and {\it associative}.
Indeed, if the term with $d\lambda$ is omitted at the r.h.s. of (\ref{47}),
there remain $3g-2$ adjusting parameters -- exactly as many as
there are holomorphic $\sigma$-odd 3-differentials ${P_k(\lambda)d\lambda\over
y^3}$,
$k=0,\ldots,3g-3$ that can arise at the l.h.s. in the triple products of
$dW_I$. In other words the algebra in hyperelliptic case is obviously
associative since it is isomorphic to the ring of polynomials
$\{ P_k(\lambda) \}$ -- multiplication of all polynomials (of one variable)
modulo some ideal.

The same counting can be easily repeated for the general surface with $n$
punctures\footnote{The simplest way to do this is to obtain all the
punctures degenerating handles of the Riemann surface of higher genus.} and
leads to the conclusion that the closed associative algebra does always exist
for the hyperelliptic case.


According to sect.\ref{eqs} the WDVV eqs are immediately
implied by the residue formula (\ref{resfor}) and associativity of the
algebra (\ref{algmult}) because the terms with $d\omega$ at the r.h.s. of
(\ref{algmult}) do not contribute into (\ref{resfor}).

\section{Examples \label{Examples}}
\setcounter{equation}{0}
\subsection{Hyperelliptic curves in the Seiberg-Witten theory}
This section contains a series of examples of the algebras of 1-differentials
and WDVV equations, arising in particular ${\cal N}=2$ supersymmetric models
in four and five dimensions.
According to \cite{GKMMM} and \cite{SWI1}, models without matter
(pure gauge theories) can be described in terms of integrable systems
of the periodic Toda-chain family. The corresponding spectral curves
are hyperelliptic, at least, for the
classical simple groups $G$. The function ${\cal P}(\lambda, w)$ is nothing
but
the characteristic polynomial of the dual affine algebra ${\hat G}^{\vee}$
and looks like (parameter $s$ was introduced in (\ref{adjot}))
\be
{\cal P}(\lambda, w) = 2P(\lambda) - w - \frac{Q_0(\lambda)}{w},
\ \ \ \ d\omega={dw\over w},\ \ \ \ Q_0(\lambda)=\lambda^{2s}
\label{curven}
\ee
while
\be
dS = \lambda \frac{dw}{w}
\label{dSn}
\ee
Here $P(\lambda)$ is characteristic polynomial of the
algebra $G$ itself, i.e.
\be
P(\lambda) = \det(G - \lambda I) =
\prod_i (\lambda - \lambda_i)
\ee
where determinant is taken in the first fundamental representation
and $\lambda_i$'s are the eigenvalues of the algebraic element $G$.
For classical simple algebras we have:
\be
A_n:\ \ \ P(\lambda) = \prod_{i=1}^{N_c}(\lambda - \lambda_i), \ \ \
N_c = n+1,\ \ \ \sum_{i=1}^{N_c} \lambda_i = 0; \ \ \
s=0; \nn \\
B_n:\ \ \ P(\lambda) = \lambda\prod_{i=1}^n(\lambda^2 - \lambda_i^2);
\ \ \ s=2 \nn \\
C_n:\ \ \ P(\lambda) = \prod_{i=1}^n(\lambda^2 - \lambda_i^2);
\ \ \ s=-2 \nn \\
D_n:\ \ \ P(\lambda) = \prod_{i=1}^n(\lambda^2 - \lambda_i^2);
\ \ \ s=2
\label{charpo}
\ee
For exceptional groups, the curves arising as the characteristic polynomial
of the dual affine algebras do not acquire the hyperelliptic form -- and
they are left beyond the scope of this paper (though they can still have
enough symmetries to suit into the general theory of s.4).

The list (\ref{charpo}) implies that one needs to analyze
separately the cases of $A_n$ and the other three series.

Above formulas are well adjusted for generalizations.
In particular, in order to include massive hypermultiplets in the
first fundamental representation one can just change
$Q_0(\lambda)$ for $Q(\lambda) = Q_0(\lambda)\prod_{\iota = 1}^{N_f}
(\lambda - m_\iota)$ if $G=A_n$ \cite{HO} and $Q(\lambda) = Q_0(\lambda)
\prod_{\iota = 1}^{N_f}(\lambda^2 - m^2_\iota)$ if $G=B_n,C_n,D_n$ \cite{AS}
(sometimes it can also make sense
to shift $P(\lambda)$ by an $m$-dependent polynomial $R(\lambda)$, which does
not depend on $\lambda_i$).

These cases of the matter hypermultiplets included can be described by the
periodic $XXX$-chain family \cite{SWI2}.

Working with
these models it is sometimes convenient to
change the $w$-variable, $w \ \rightarrow \ \sqrt{Q(\lambda)}w$
so that the curve ${\cal P}(\lambda, w) = 0$ turns into
\be
w + \frac{1}{w}  = 2\frac{P(\lambda) + R(\lambda)}{\sqrt{Q(\lambda)}}
\label{curven2}
\ee

Another possible generalization is the extension
to $5d$ models \cite{N}. Then it is enough to change only the
formula (\ref{dSn}),
\be
dS^{(5)} = \log\lambda\ \frac{dw}{w}
\label{dS5}
\ee
and put $Q_0=\lambda^{N_c/2}$ while eq.(\ref{curven}) remains intact.
From the point of view of integrable systems this corresponds to changing
the Toda chain for the relativistic Toda chain.

Examples that do not suit into this scheme are related to the
UV-finite models, of which the simplest one is the
$SU(N_c)$ gauge theory in $4d$ with massive hypermultiplet in the
adjoint representation. The corresponding integrable system
is the elliptic Calogero model, and our analysis in s.\ref{Calogero}
reveals that it does not satisfy WDVV equations: the closed algebra
of holomorphic differentials still exists, but is not
associative -- in accordance with the general reasoning in s.4.

To conclude these short remarks, let us note that, as we shall see below,
associativity of algebras of the holomorphic 1-differentials on
hyperelliptic curve reduces afterwards to associativity of the polynomial
algebras. Therefore, we consider here the reference pattern of such an
algebra. Namely, we look at the algebra of $N$ one-variable polynomials
$p_{i}(\lambda)$ of degree $N-1$ whose product defined by modulo
of a polynomial $P'$ of degree $N$. More concretely, the algebra
is defined as
\be\label{polass}
p_i(\lambda)p_j(\lambda)=C^k_{ij}p_k(\lambda)W(\lambda)+\xi_{ij}(\lambda)
P'(\lambda)
\ee
where $W(\lambda)$ is some fixed linear combination of $p_i(\lambda)$:
$W(\lambda)\equiv \sum q_lp_l(\lambda)$
co-prime with $P'(\lambda)$. Then, the l.h.s. of this relation is a
polynomial of degree $2(N-1)$ with $2N-1$ coefficients which are to be
adjusted by the free parameters in the r.h.s. Since $\xi_{ij}$ is of degree
$N-2$, there are exactly $N+(N-1)=2N-1$ free parameters. The same counting is
correct for the triple products of the polynomials and the algebra
is associative, since the condition $P'(\lambda)=0$ leads to the
factor-algebra over the ideal $P'(\lambda)$. The same argument does not work
for the algebra of holomorphic 1-differentials, since their product belongs
to a different space (of quadratic differentials) -- see \cite{MMM2}.

\subsection{Perturbative limit of the pure gauge theories}
Now we discuss the simplest case of curves corresponding to
the perturbative formulas of sect.3.
Perturbative limit of the curve (\ref{curven2}) is
\be\label{pertcurv}
w = 2\frac{P(\lambda)}{\sqrt{Q(\lambda)}}
\ee
and the corresponding
\be\label{pertdS}
dS^{(4)} = \lambda d\log\left(\frac{P(\lambda)}{\sqrt{Q(\lambda)}}\right);
\nn \\
dS^{(5)} = \log\lambda\  d\log\left(\frac{P(\lambda)}{\sqrt{Q(\lambda)}}\right)
\ee
We consider first the simplest case of the pure gauge $SU(N_c)$ Yang-Mills
theory, i.e. $Q(\lambda)=Q_0(\lambda)=1$.
The both cases of $4d$ and $5d$ theories
can be treated simultaneously. Indeed, they are described by the same Riemann
surface (\ref{pertcurv}) which is nothing but
the sphere with punctures\footnote{These punctures emerge as a
degeneration of the handles of the hyperelliptic surface so that the
$a$-cycles encircle the punctures.}, the only difference being the
restrictions imposed onto the set of moduli $\{\lambda_i\}$'s --
in the $4d$ case
they are constrained to satisfy the condition $\sum_i^{N_c} \lambda_i=0$,
while in
the $5d$ one -- $\prod_i^{N_c} \lambda_i=1$.
Now the set of the $N_c-1$ independent
canonical holomorphic differentials is\footnote{It can be instructive to
describe in more details why (\ref{difff}) arises not only in $4d$, but also
in $5d$ models. Strictly speaking, in the $5d$ case, one should consider
differentials on annulus, not on sphere. Therefore, instead of
${d\lambda\over\lambda-\lambda_i}$, one rather needs to take
$\displaystyle{\sum_{m=-\infty}^{+\infty}{da\over
a-a_i+2n}\sim\coth{a-a_i\over 2}{da\over
2}={\lambda+\lambda_i\over\lambda-\lambda_i}{d\lambda\over 2\lambda}}$, where
$\lambda\equiv e^a$. Considering now, instead of
$\displaystyle{d\omega_i={d\lambda\over
\lambda-\lambda_i}-{d\lambda\over\lambda-\lambda_{N_c}}={\lambda_{iN_c}
d\lambda\over (\lambda-\lambda_i)(\lambda-\lambda_{N_c}}}$, differentials
$\displaystyle{d\omega_i={\lambda+\lambda_i\over\lambda-\lambda_i}
{d\lambda\over 2\lambda} -{\lambda+\lambda_{N_c}\over\lambda-\lambda_{N_c}}
{d\lambda\over
2\lambda}= {\lambda_{iN_c}d\lambda\over
(\lambda-\lambda_i)(\lambda-\lambda_{N_c})}}$, we obtain exactly formula
(\ref{difff}).}

\be\label{difff}
d\omega_i=\left({1\over\lambda-\lambda
_i}-{1\over\lambda -\lambda_{N_c}}\right)d\lambda=
{\lambda_{iN_c}d\lambda\over
(\lambda -\lambda_i)(\lambda -\lambda_{N_c})},\ \ \ i=1,...,N_c,
\ \ \ \lambda_{ij}\equiv \lambda_i-\lambda_j
\ee
and one can easily check that $\lambda_i=a_i$ ($a$-periods) in the $4d$ case
and $\lambda_i=e^{a_i}$ in the $5d$ case, i.e. $\sum_i a_i=0$.

The next step is to expand the product of two such differentials
\be
d\omega_id\omega_j=\left(C_{ij}^kd\omega_k\right)\left(q_ld\omega_l\right)+
\left(D^k_{ij}d\omega_k\right)d\log P(\lambda),\ \ \ d\log P(\lambda)=
\sum_l^{N_c}{d\lambda\over\lambda -\lambda_l}
\ee
Converting this expression to the normal form with
common denominator $P(\lambda)$,
one reproduces (\ref{polass}) for polynomial algebra.
Therefore, the structure constants $C_{ij}^k$
form the closed associative algebra and, in
accordance with our general formulas, can be expressed through the
prepotential by residue formulas
\be
F_{ijk}=\stackreb{d\log P=0}{\res} {d\omega_id\omega_jd\omega_k\over d\log
P(\lambda)d\lambda} \ \ \ \hbox{for the $4d$ case},\\
F_{ijk}=\stackreb{d\log P=0}{\res} {d\omega_id\omega_jd\omega_k\over d\log
P(\lambda){d\lambda\over\lambda}} \ \ \ \hbox{for the $5d$ case}
\ee
These residues can be easily calculated. The only technical trick is to
calculate the residues not at zeroes of $d\log P$ but at poles of
$d\omega$'s. This can be done immediately since there are no contributions
from the infinity $\lambda=\infty$.
The final results have the following form
\be
\hbox{\underline{for the $4d$ case}}:\\
F_{iii}=\sum_{k\ne i}{1\over a_{ik}}+{6\over a_{iN_c}}+
\sum_{k\ne N_c}{1\over a_{kN_c}},
\\
F_{iij}= {3\over a_{iN_c}}+{2\over a_{jN_c}}+\sum_{k\ne i,j,N_c}
{1\over a_{kN_c}}-{a_{jN_c}\over a_{iN_c}a_{ij}},
\ \ \ i\ne j,\\
F_{ijk}=2\sum_{l\ne N_c}{1\over a_{lN_c}}-\sum_{l\ne i,j,k,N_c}
{1\over a_{lN_c}},
\ \ \ i\ne j\ne k;
\ee

\be
\hbox{\underline{for the $5d$ case}}:\\
2F_{iii}=\sum_{k\ne i}\coth a_{ik}+
6\coth a_{iN_c}+\sum_{k\ne N_c}\coth
 a_{kN_c},\\
2F_{iij}= -\coth a_{ij}+
4\coth a_{iN_c}+2\coth a_{jN_c}+\sum_{k\ne
i,j,N_c}\coth a_{kN_c}+N_c,\ \ \ i\ne j,\\
2F_{ijk}= 2\sum_{l\ne N_c}\coth a_{lN_c}-\sum_{l\ne i,j,k,N_c}\coth
 a_{lN_c}+N_c,\ \ \ i\ne j\ne k
\ee
Now it is immediate
to check that the prepotential in the $4d$ case really is given by the
formula (\ref{FA}), while in the $5d$ case -- by (\ref{FARTC}).
These results are in perfect agreement with the results of s.3.

Now let us turn to the case of massive hypermultiplets included and, for the
sake of simplicity, restrict ourselves to the $4d$ case. Then, there
arise additional differentials corresponding to the derivatives of $dS$
(\ref{pertdS}) with respect to masses. They are of the form
\be\label{massdiff}
dW_{\iota}=-{1\over 2}{d\lambda\over \lambda -\lambda_{\iota}}
\ee
Again, full set of the differentials gives rise to
the closed associative algebra.
As earlier, this can be proved by reducing the system of differentials to the
polynomial one (this time the common denominator should be
$P(\lambda)Q(\lambda)$). Then, applying the residue formula, one reproduces
formula (\ref{26}) for the perturbative prepotential with $r=1$. In order to
obtain (\ref{26}) with arbitrary coefficient $r$, one needs to consider
instead of $\sqrt{Q(\lambda)}$ in (\ref{pertcurv})-(\ref{pertdS})
$Q(\lambda)^{r/2}$. This only changes normalization of the differentials
(\ref{massdiff}):  $dW_{\iota}=-{r\over 2}{d\lambda\over \lambda -
\lambda_{\iota}}$. Now it is straightforward
to use the residue formula\footnote{Let us note that, despite
the differentials
$dW_{\iota}$ have the pole at infinity, this does not contribute
into the residue formula because of the quadratic pole of $d\lambda$ in the
denominator.}

\be\label{residueQ}
F_{IJK}=\stackreb{d\log {P\over\sqrt{Q}}=0}{\res}
{dW_IdW_JdW_K\over d\log {P(\lambda)\over\sqrt{Q(\lambda)}}d\lambda},
\ \ \ \ \
\left\{I,J,K,\dots\right\}=\left\{i,j,k,\ldots|\iota,\iota',
\iota'',\ldots\right\}
\ee
and obtain
\be\label{perth}
F_{iii}=\sum_{k\ne i}{1\over \lambda_{ik}}+{6\over
\lambda_{iN_c}}+ \sum_{k\ne N_c}{1\over
\lambda_{kN_c}}-\r\sum_{\iota}{\lambda_{iN_c}\over
\lambda_{i{\iota}}\lambda_{{\iota}N_c}},
\\
F_{iij}= {3\over \lambda_{iN_c}}+{2\over \lambda_{jN_c}}+\sum_{k\ne i,j,N_c}
{1\over \lambda_{kN_c}}-
{\lambda_{jN_c}\over \lambda_{iN_c}\lambda_{ij}}-\r\sum_{\iota}{1\over
\lambda_{{\iota}N_c}},
\ \ \ i\ne j,\\
F_{ijk}=2\sum_{l\ne N_c}{1\over \lambda_{lN_c}}-\sum_{l\ne i,j,k,N_c}
{1\over \lambda_{lN_c}}-\r\sum_{\iota}{1\over\lambda_{{\iota}N_c}},
\ \ \ i\ne j\ne k;\\
F_{ii{\iota}}= \r\left({\f\lambda_{i{\iota}}}-{\f\lambda_{{\iota}N_c}}
\right);\\
F_{ij{\iota}}=-\r{\f\lambda_{{\iota}N_c}},\ \ \ i\ne j;\\
F_{i{\iota}{\iota}}=-\r{\lambda_{iN_c}\over\lambda_{i{\iota}}
\lambda_{{\iota}N_c}};\\
F_{{\iota}{\iota}{\iota}}=
\r\sum_i{\f\lambda_{{\iota}i}}+{r^2\over 4}\sum_{{\iota}'}
{\f\lambda_{{\iota}{\iota}'}};\\
F_{{\iota}{\iota}{\iota}'}=
-{r^2\over 4}{\f \lambda_{{\iota}{\iota}'}},\ \ \ {\iota}\ne {\iota}';\\
F_{i{\iota}{\iota}'}=F_{{\iota}{\iota}'{\iota}''}=0
\ee
These formulas immediately lead to the prepotential (\ref{26}) upon the
identification $\lambda_i=a_i$ and $\lambda_{\iota}=m_{\iota}$.

At last, let us consider the case of other classical groups. We also take into
account the massive hypermultiplets. Let us also note that, in this case, the
curve is invariant w.r.t. the involution $\rho: \lambda\to
-\lambda$. Among all the holomorphic differentials there is a subset of the
$\rho$-odd ones that can be obtained by differentiating the $\rho$-odd
generating differential $dS\cong \lambda\left(\sum_i {2\lambda d\lambda\over
\lambda^2-a_i^2}-r\sum_{\iota} {\lambda d\lambda\over \lambda^2-m_{\iota}^2}-
s{d\lambda\over\lambda}\right)$ w.r.t. the moduli:
\be
dW_i={2a_i d\lambda\over \lambda^2-a_i^2},\ \ \ \
dW_{\iota}=-{r m_{\iota} d\lambda\over \lambda^2-m_{\iota}^2}
\ee
Look at the algebra of these differentials:
\be\label{ch}
{d\lambda\over \lambda^2-\lambda_I^2} {d\lambda\over \lambda^2-\lambda_J^2}
\sim C_{IJ}^K {d\lambda\over \lambda^2-\lambda_K^2} dW+
\xi_{IJ}(\lambda)
\left(\sum_i {2\lambda d\lambda\over \lambda^2-a_i^2}-r\sum_{\iota}
{\lambda d\lambda\over \lambda^2-m_{\iota}^2}-s{d\lambda\over\lambda}\right)
\ee
To understand that this algebra is closed and associative, it is sufficient
to note that the l.h.s. of (\ref{ch}) is a linear combination of the
differentials ${(d\lambda)^2\over \lambda^2-\lambda_I^2}$ and ${(d\lambda)^2
\over (\lambda^2-\lambda_I^2)^2}$ -- totally $2N\equiv 2n+2N_f$ independent
quadratic $\rho$-even differentials. However, one should also take into
account the condition which cancels the pole at infinity of
${(d\lambda)^2\over \lambda^2-\lambda_I^2}$. Thus, the number of
{\it holomorphic} $\rho$-even quadratic differentials is equal to $2N-1$.
This quantity perfectly suits the $2N-1$ adjusting parameters in the r.h.s.
of (\ref{ch}) -- $N$ $\rho$-odd differentials in the first term and $N-1$
$\rho$-even differentials in the second term. Indeed, even differentials are
${\lambda d\lambda\over \lambda^2-\lambda_I^2}$, i.e. have the pole at
infinity. The same does the
${dw\over w}=\left(\sum_i {2\lambda d\lambda\over \lambda^2-a_i^2}-\sum_{\iota}
{r\lambda d\lambda\over \lambda^2-m_{\iota}^2}-s{d\lambda\over\lambda}\right)$.
However, since all these quantities can be rewritten as depending on
$\lambda^2$, the cancellation of the pole of the differentials leads to
their large-$\lambda$ behaviour ${d\lambda\over\lambda^3}$ that
simultaneously cancels the pole of ${dw\over w}$. Thus, there is exactly one
condition imposed onto $N$ $rho$-even differentials.

The associativity is checked in complete analogy with the above
consideration -- there are $3N-3$ holomorphic $\rho$-odd cubic differentials
and exactly $N+(2N-3)=3N-3$ adjusting parameters.
Thus, the algebra (\ref{ch}) is really closed associative\footnote{In sect.5.5
we show how this algebra reduces to the
polynomial one.}. The application of the residue formula (\ref{residueQ})
gives now (\ref{adjot}).

Now we turn to the less trivial examples corresponding to full hyperelliptic
curves.

\subsection{Pure gauge model (Toda chain system)}
This example was analyzed in detail in ref.\cite{MMM}.

The spectral curve has the form
\be\label{specTC}
{\cal P}(\lambda,w)=2P(\lambda)-w-{1\over w},\ \ \
P(\lambda ) \equiv \sum^{N_c}_{k=0} s_k\lambda ^k
\ee
and can be represented in the standard hyperelliptic form after the change of
variables $Y = \frac{1}{2}\left(w - \frac{1}{w}\right)$
\be
Y^2 = P^2(\lambda ) - 1
\ee
and is of genus $g = N_c-1$
\footnote{One can take for the $g$ moduli
the set $\{h_k\}$ or instead the set of periods $\{a_i\}$.
This particular family is associated with the Toda-chain
hierarchy, $N_c$ being the length of the chain.}.
Note that $s_{N_c}=1$ and that the condition of the unit determinant of the
matrix in the group $SU(N_c)$ implies here that $s_{N_c-1}=0$ (this is
equivalent to the condition $\sum a_i=0$).
The generating differential $dS$ has the form
\be\label{dSTC}
dS=\lambda{dw\over w}=\lambda{dP\over Y}
\ee

Now let us discuss associativity of the algebra of holomorphic differentials
\be
d\omega_i(\lambda )d\omega_j(\lambda ) =
C_{ij}^k d\omega_k(\lambda )
dW(\lambda ) \ {\hbox{mod}}\ {dw\over w}
\label{.}
\ee
We require that $dW(\lambda)$ and $dP(\lambda )$ are co-prime.

If the algebra (\ref{.}) exists, the structure constants
$C_{ij}^k$ satisfy the associativity condition.
But we still need to show that
it indeed exists, i.e. that once $dW$ is given, one can find
($\lambda $-independent) $C_{ij}^k$. This is a simple exercise:
all $d\omega_i$ are linear combinations of
\be\label{match}
dv_k(\lambda ) = \frac{\lambda ^{k}d\lambda }{Y}
\cong {\partial dS\over\partial s_k}, \ \ \ k=0,\ldots,g-1
\ee
where
\be
dv_k(\lambda ) = \sigma_{ki}d\omega_i(\lambda ), \ \ \
d\omega_i = (\sigma^{-1})_{ik}dv_k, \ \ \
\sigma_{ki} = \oint_{A_i}dv_k,
\label{sigmadef}
\ee
also $dW(\lambda ) = q_kdv_k(\lambda )$.
Thus, (\ref{.}), after multiplying all 1-differentials by
$Y$, reduces to the algebra of polynomials
$\left(\sigma^{-1}_{ii'}\lambda ^{i'-1}\right)$ that is as usual closed and
associative. This means that $C_{ij}^k$ satisfy the
associativity condition
\be
C_iC_j = C_j C_i
\ee
Now the residue formula has the form \cite{MMM}:
\be
F_{ijk} = \frac{\partial^3F}{\partial a_i\partial a_j
\partial a_k} = \frac{\partial F_{ij}}{\partial a_k} = \nn \\
= \stackreb{d\lambda =0}{{\res}} \frac{d\omega_id\omega_j
d\omega_k}{d\lambda\left(\frac{dw}{w}\right)} =
\stackreb{d\lambda =0}{{\res}} \frac{d\omega_id\omega_j
d\omega_k}{d\lambda\frac{dP}{Y}} =
\sum_{\alpha} \frac{\hat\omega_i(\lambda_\alpha)\hat\omega_j
(\lambda_\alpha)\hat\omega_k(\lambda_\alpha)}{P'(\lambda_\alpha)
/\hat Y(\lambda_\alpha)}
\label{v}
\ee
The sum at the r.h.s. goes over all the $2g+2$ ramification points
$\lambda_\alpha$ of the hyperelliptic curve (i.e. over the zeroes
of $Y^2 = P^2(\lambda )-1 = \prod_{\alpha=1}^N(\lambda - \lambda_\alpha)$);
\  $d\omega_i(\lambda) = (\hat\omega_i(\lambda_\alpha) +
O(\lambda-\lambda_\alpha))d\lambda$,\ $\ \ \ \hat Y^2(\lambda_\alpha) =
\prod_{\beta\neq\alpha}(\lambda_\alpha - \lambda_\beta)$.

Now following the line of sect.2, we can
define the metric:
\be
\eta_{kl}(dW) =
\stackreb{d\lambda =0}{{\res}} \frac{d\omega_kd\omega_l
dW}{d\lambda\left(\frac{dw}{w}\right)} =
\stackreb{d\lambda =0}{{\res}} \frac{d\omega_kd\omega_l
dW}{d\lambda\frac{dP}{Y}} = \\ =
\sum_{\alpha} \frac{\hat\omega_k(\lambda_\alpha)\hat\omega_l
(\lambda_\alpha)\hat W(\lambda_\alpha)}{P'(\lambda_\alpha)
/\hat Y(\lambda_\alpha)}
\label{vv}
\ee
In particular, for $dW = d\omega_k$ (which is evidently co-prime with
${dP\over Y}$), $\eta_{ij}(d\omega_k) = F_{ijk}$ -- this choice immediately
give rise to the WDVV in the form (5) of paper \cite{MMM}.

Given (\ref{.}), (\ref{v}) and (\ref{vv}), one can check that
\be
F_{ijk} = \eta_{kl}(dW)C_{ij}^k(dW).
\label{vvv}
\ee
Note that $F_{ijk} = {\partial^3F\over\partial a_i\partial a_j
\partial a_k}$ at the l.h.s. of (\ref{vvv}) is independent
of $dW$! The r.h.s. of (\ref{vvv}) is equal to:
\be
\eta_{kl}(dW)C_{ij}^l(dW) =
\stackreb{d\lambda =0}{{\res}} \frac{d\omega_kd\omega_l
dW}{d\lambda\left(\frac{dw}{w}\right)} C_{ij}^l(dW)
\stackrel{(\ref{.})}{=} \\ =
\stackreb{d\lambda =0}{{\res}} \frac{d\omega_k}
{d\lambda\left(\frac{dw}{w}\right)}
\left(d\omega_id\omega_j - p_{ij}\frac{dPd\lambda}{Y^2}\right) =
F_{ijk} - \stackreb{d\lambda =0}{{\res}} \frac{d\omega_k}
{d\lambda\left(\frac{dP}{Y}\right)}p_{ij}(\lambda)\frac{dPd\lambda}{Y^2}
= \\ = F_{ijk} - \stackreb{d\lambda =0}{{\res}}
\frac{p_{ij}(\lambda )d\omega_k(\lambda)} {Y}
\ee
It remains to prove
that the last item is indeed vanishing for any $i,j,k$.
This follows from the
fact that $\frac{p_{ij}(\lambda )d\omega_k(\lambda )}{Y}$
is singular only at zeroes of $Y$, it is not singular at
$\lambda =\infty$ because
$p_{ij}(\lambda)$ is a polynomial of low enough degree
$g-2 < g+1$. Thus the sum of its residues at ramification points
is thus the sum over {\it all} the residues and therefore vanishes.

This completes the proof of associativity condition for any $dW$.

\subsection{$5d$ pure gauge theory (Relativistic Toda chain)}
The above proof can be almost literally transferred to the case of the
relativistic Toda chain system corresponding to the $5d$ $N=2$ SUSY pure gauge
model with one compactified dimension \cite{N}. The main reason is
that in the case of relativistic Toda chain the spectral curve is a minor
modification of (\ref{specTC}) having the form \cite[eq.(2.23)]{N}
\be\label{specRTC}
w + {1\over w} = \left(\zeta\lambda\right)^{-N_c/2}P(\lambda ),
\ee
which can be again rewritten as a {\em hyperelliptic} curve in terms of the
new variable $Y\equiv \left(\zeta\lambda\right)^{N_c/2}\left(w-
{1\over w}\right)$
\be
Y^2 = P^2(\lambda ) - 4\zeta^{2N_c}\lambda^{N_c}
\ee
where $\lambda \equiv e^{2\xi}$, $\xi $ is the "true" spectral
parameter of the relativistic Toda chain and $\zeta$ is its coupling constant.

The difference with the $4d$ case is at the following two points.
The first is that this time $s_0\sim \prod e^{a_i}=1$
\footnote{This formula looks quite natural since the relativistic Toda chain
is a sort of "group generalization" of the ordinary Toda chain.} but instead
$s_{N_c-1}$ no longer vanishes and becomes a modulus.

The second new point is that the generating differential instead of
(\ref{dSTC}) is now
\be\label{dSRTC}
dS^{(5)} = \xi{dw\over w} \sim \log\lambda {dw\over w}
\ee
so that its periods are completely different from those of $dS^{(4)}$.
However, the set of holomorphic differentials
\be
dv_k = {\partial dS^{(5)}
\over\partial s_k} \cong {\lambda^{k-1}d\lambda\over Y},
\ \ \ k=1,...,g
\ee
literally coincides with the set (\ref{match}), because $k$ now runs through
$\ 1,\ldots,g\ $ instead of $\ 0,\ldots,g-1\ $ but there is the extra factor of
$\lambda$ in the denominator. Thus, the algebra of differentials remains just
the same closed associative algebra, but the residue formula is a little
modified.

Indeed, when passing from the first line of (\ref{resfor}) to the second
line, there will arise the extra factor $\lambda$ due to the shift of the
index of differentials $dv_k$ by 1 in the $5d$ case as compared with the $4d$
one. Thus, finally the residue formula acquires the following form
\be
F_{ijk} =
\stackreb{d\lambda=0}{{\res}} \frac{d\omega_id\omega_j
d\omega_k}{\left(\frac{d\lambda}{\lambda}\right)\left(\frac{dw}{w}\right)} =
\stackreb{d\lambda=0}{{\res}} \frac{d\omega_id\omega_j
d\omega_k}{\left(\frac{d\lambda}{\lambda}\right)\left(\frac{dP}{Y}\right)} =
\sum_{\alpha} \lambda_{\alpha}\frac{\hat\omega_i(\lambda_\alpha)\hat\omega_j
(\lambda_\alpha)\hat\omega_k(\lambda_\alpha)}{P'(\lambda_\alpha)
/\hat Y(\lambda_\alpha)}
\ee

\subsection{Pure gauge theory with $B_n$, $C_n$ and $D_n$ gauge groups}
Now we consider shortly the case of the pure gauge theory with the groups
$G\ne A_n$. This example is of importance, since it demonstrates that, in the
cases when naively there is no closed associative algebra of holomorphic
1-differentials, it really exists due to some additional symmetries of the
curve.

The curve in this case has the form (\ref{curven}) with $Q_0(\lambda)
=\lambda^{2s}$ as in
(\ref{charpo}), while the generating differential -- (\ref{dSn}). After
change of variables
\be
2Y(\lambda)=w-{Q_0(\lambda)\over w}
\ee
the curve acquires the standard hyperelliptic form
\be\label{hra}
Y^2=P^2(\lambda)-Q_0(\lambda)
\ee
Naively, we can repeat the procedure applied in the previous examples. The
new problem, however, is that the genus of this curve now is $2n-1$ (see
(\ref{charpo})), while there are only $n$ moduli, and, therefore, $n$
holomorphic differentials which can be obtained from the generating
differential (\ref{dSn}). Therefore, one might expect that the algebra formed
by these differentials is too small to be closed (and associative).

However, it turns out that, due to the additional $Z_2$-symmetry, this is not
the case. Indeed, let us note that the curve (\ref{hra}) admits the
involution $\rho:\lambda\to -\lambda$ (in addition to that $Y\to -Y$).
Therefore, in the complete set of holomorphic $\sigma$-odd
differentials on the
hyperelliptic curve (\ref{hra}) (see sect.3.4)
$dv_i={\lambda^i d\lambda\over Y}$,
$i=0,\ldots,2n-2$, there are $n-1$ $\rho$-even and $n-1$ $\rho$-odd
differentials. Keeping in mind the rational examples of sect.5.2, we should
look at the subalgebra formed by the $\rho$-odd differentials that only can be
obtained from the generating differential $dS$, i.e. at $dv_i$ with $i=2k$ --
or at the canonical $\rho$-odd differentials $d\omega^o_{i}$
\be\label{pred}
d\omega^o_id\omega^o_j=C^k_{ij}d\omega^o_kdW\ \ \hbox{mod}\;{dw\over w}
\ee
Since ${dw\over w}$ is $\rho$-even, it can be multiplied only by $\rho$-even
holomorphic differentials. On the other hand, there are $2n-2$ $\rho$-even
quadratic holomorphic differentials. This perfectly matches the number of the
free adjusting parameters at the r.h.s. of {\ref{pred}), i.e. algebra
(\ref{pred}) is really closed (and associative, as one can obtain by the
analogous counting).

The fact of closeness and associativity of this
algebra can be demonstrated in
another and easier way -- like it has been done in our previous examples.
Indeed, let us multiply all the differentials by $\lambda Y$ that is the
common denominator (since ${dw\over w}={dP\over Y}+
s\left(1-{P\over Y}\right){d\lambda\over \lambda}$). Then, the algebra is
reduced (as usual in the hyperelliptic case) to the polynomial algebra. More
precisely, this is the closed associative algebra of $n-1$ odd polynomials
$p_i^o$ of degree $2n-3$ factorized over the ideal which is an even polynomial
$\lambda P'(\lambda)$ of the degree $2n$:
\be
p_i^o(\lambda)p_j^o(\lambda)=C^k_{ij}p_k^o(\lambda)W(\lambda)
+\xi_{ij}^e(\lambda)\lambda P'(\lambda)
\ee
where $\xi_{ij}^e$ is some even polynomial. The degree of this expression is
$4n-6$, at the l.h.s. of it there are $2n-3$ independent coefficients of the
even polynomial of degree $4n-6$ with no constant term, while at the r.h.s.
there are $n-1$ adjusting parameters $C_{ij}^k$ and $n-2$ adjusting
coefficients of the polynomial $\xi^e_{ij}(\lambda)$, totally $2n-3$ free
parameters. This counting proves that the described algebra is closed (and
associative as polynomial algebra).

\subsection{Hypermultiplets in the fundamental representation
(Spin chains)}
To conclude the section, we consider the example of hyperelliptic curve
associated with the $SU(N_c)$ Yang-Mills theory with mass
hypermultiplets.

The set of holomorphic 1-differentials $\{d\omega_i\}$ relevant
in the Toda chain case (s.5.3 and \cite{MMM}) should
be now enlarged to include meromorphic ones, with the simple poles at
the points $m_{\iota}^\pm$. The spectral curve is given by:
\be\label{qcd}
{\cal P}(\lambda,w) = 2P(\lambda) -
w - \frac{Q(\lambda)}{w}
\ee
where
\be
P(\lambda) = \sum_{k=0}^{N_c} s_k\lambda^k + R(\lambda) \ \ \ \ \ \
Q(\lambda) = \prod_{s=1}^{N_f} (\lambda - m_{\iota})
\ee
$s_k$ are the Schur polynomials of the Hamiltonians $h_k$ and do not
depend on $m_{\iota}$, while
$R(\lambda)$ is instead an $h_k$-independent polynomial in $\lambda$
of degree $N_c-1$,  which is non-vanishing only for $N_f>N_c$
\cite{HO} (see also \cite{SWI2}).
It is convenient to introduce another function $Y$
\be
2Y = w - \frac{Q(\lambda)}{w}, \ \ \ \ \
Y^2 = P^2(\lambda) - Q(\lambda) = \prod(\lambda - \lambda_\alpha)
\ee
The generating differential
\be\label{dS1}
dS = \lambda\frac{dw}{w} = \lambda\left(
\frac{dP}{Y} - \frac{PdQ}{2QY} + \frac{dQ}{2Q}\right)
\ee
satisfies the defining property
\be
\frac{\partial dS}{\partial a_k} = d\omega_k\ \
\left(\{ dW_a\} = \frac{Pol}{QY}d\lambda\right)
\ee
and the algebra of multiplication acquires the form:
\be
dW_a dW_b = C_{ab}^c dW_cdW \ {\hbox{mod}}\ \frac{dw}{w}
\ee
$a = 1,\ldots,g+ N_f = \{i,s\}$, $g = N_c-1$. The general formulas in the
particular case (\ref{qcd}) are
\be
{\partial dS\over \partial a_I} \cong
{{\partial P\over\partial a_I}- {1\over 2w}{\partial Q\over\partial a_I}
\over {P'-{Q'\over 2w}}}{dw\over w}
\equiv {\cal T}_I{dw\over {\cal P}'w}
\ee
and
\be
{dw\over w} = {dP\over Y} - {dQ\over 2Y(P+Y)}
\nn \\
\left.Y{dw\over w}\right|_{Y=0} = dP - {dQ\over 2P}
\ee
The calculations (along the lines of \cite{MMM}) lead to the same conclusion:
\be
a_k = \oint_{A_k} dS, \ \ \ \
\oint \frac{\partial\ dS}{\partial s_l} = \oint
\frac{\lambda^ld\lambda}{Y}
\ee
\be
\frac{\partial F_{ij}}{\partial
 a_k} =
\frac{\partial F_{ij}}{\partial \lambda_\alpha}
\frac{\partial\lambda_\alpha}{\partial s_l}
\frac{\partial s_l}{\partial a_k}
\ee
The first derivative is (as usual for any hyperelliptic curve)
equal to
$\hat\omega_i\hat\omega_j(\lambda_\alpha)$, the
last one -- to $\sigma^{-1}_{kl}$. In order to find the
middle one, consider
\be
\left.\frac{\partial}{\partial s_l}
\left(P^2 - Q = \prod_\beta (\lambda - \lambda_\beta)\right)
\right|_{\lambda = \lambda_\alpha}
\ee
It gives
\be
\left. 2P\frac{\partial P}{\partial s_l} =
(2PP' - Q')\frac{\partial \lambda_\alpha}{\partial s_l}
\right|_{\lambda = \lambda_\alpha}
\ee
i.e.
\be
\frac{\partial\lambda_\alpha}{\partial s_l} =   \left.
\frac{\lambda^ld\lambda}{dP - \frac{1}{2}\frac{dQ}{P}}
\right|_{\lambda = \lambda_\alpha}
\ee
Since at $\lambda = \lambda_\alpha$ $Y=0$, we obtain:
\be
\frac{\partial\lambda_\alpha}{\partial s_l} =
  \left.\frac{\lambda_\alpha^ld\lambda}{Y\frac{dw}{w}}
\right|_{\lambda = \lambda_\alpha} =
\left.\sigma_{lk}\frac{d\omega_k}{\frac{dw}{w}}
\right|_{\lambda = \lambda_\alpha}
\ee
Putting this all together we obtain:
\be
\frac{\partial F_{ij}}{\partial a_k} =
-\stackreb{d\lambda = 0}{\res}
\frac{d\omega_id\omega_jd\omega_k}{d\lambda\frac{dw}{w}} =
\stackreb{d\omega = 0}{\res}
\frac{d\omega_id\omega_jd\omega_k}{d\lambda\frac{dw}{w}}
\ee
Since differentials in the numerator have no poles, the
integration contour that goes around the zeroes of $d\lambda$,
can be also considered as encircling the zeroes of
$d\omega = \frac{dw}{w}$ -- and we reproduce the general
result (\ref{resfor}).

The similar identity can be proved for the expressions containing
one derivative w.r.t. the remaining moduli $m_{\iota} \sim \oint dS$,
as we can still work with the period matrix. More concretely, since
\be\label{dW1}
dW_{\iota} \cong  \frac{\partial dS}{\partial m_{\iota}} \cong -
\frac{d\lambda}{w}\frac{\partial w}{\partial m_{\iota}} = -
\frac{d\lambda}{Y}\left(\frac{Q}{2(P+Y)(\lambda - m_{\iota})} +
\frac{\partial R}{\partial m_{\iota}}\right)
\ee
the derivative of ramification point is
\be
\frac{\partial \lambda_\alpha}{\partial m_{\iota}} =
-\left. \frac{Q}{2P}\frac{d\lambda}{Y\frac{dw}{w}}
\left(\frac{1}{\lambda - m_{\iota}} +
\frac{2P\frac{\partial R}{\partial m_{\iota}}}{Q}\right)\right|_{\lambda =
\lambda_\alpha} =-
\left. \frac{d\lambda}{Y\frac{dw}{w}}
\left(\frac{Q}{2(P+Y)(\lambda - m_{\iota})} +
\frac{\partial R}{\partial m_{\iota}}\right)\right|_{\lambda =
\lambda_\alpha}
=\left.\frac{dW_{\iota}}{\frac{dw}{w}}
\right|_{\lambda = \lambda_\alpha}
\ee
Thus
\be
\frac{\partial F_{ij}}{\partial m_{\iota}} =
\frac{\partial F_{ij}}{\partial \lambda_\alpha}
\frac{\partial\lambda_\alpha}{\partial m_{\iota}} =
-\stackreb{d\lambda = 0}{\res}
\frac{d\omega_i d\omega_j dW_{\iota}}{d\lambda \frac{dw}{w}}
= \stackreb{\frac{dw}{w} = 0}{\res}
\frac{d\omega_i d\omega_j dW_{\iota}}{d\lambda \frac{dw}{w}}
\ee
again in accordance with (\ref{resfor}).
Note, that according to our agreements the punctures $m_\iota$
are included into the set of zeroes of $d\lambda$ (along with
ramification points $\lambda_\alpha$).

Double and triple-derivatives of ${\cal F}$ with respect to the
masses $m_\iota$ can not be obtained from the ordinary
period matrices of the hyperelliptic curves. They can be only
evaluated by the general method of s.4. This calculation,
which is the subject of the next subsection s.\ref{rfmd},
helps also to illustrate the possible subtleties of the general proof.

Let us note that there are two natural choices of the generating differential
$dS$ for the case of theory with fundamental matter. One choice is given by
formula (\ref{dS1}) used throughout this subsection. Another one is
given by $dS=\lambda{d\hat w\over\hat w}=\lambda d\log{P-Y\over P+Y}$,
where $\hat w$ is defined in another parameterization of the curve $\hat w
+{1\over\hat w}={P\over\sqrt{Q}}$.  The latter choice is
usually accepted in the literature \cite{SW2,SWI2,HO,AS}, and has been used in
the course of our perturbative consideration of sect.5.2. Note that the
differentials (\ref{dW1})
\be
dW_{\iota}^{(1)} \cong
-\frac{d\lambda}{Y}\left(\frac{P-Y}{2(\lambda - m_{\iota})} +
\frac{\partial R}{\partial m_{\iota}}\right)=
dW^{(2)}_{\iota}+{1\over 2}{d\lambda\over\lambda -m_{\iota}}
\ee
have poles at $m_{\iota}$ on one of the sheets, while the differentials
$dW^{(2)}_{\iota}$ associated with the other $dS$
possess the poles on both sheets.
However, these two choices give rise to the identical results.

\subsection{Residue formula for mass differentials \label{rfmd}}

To make our consideration of concrete examples exhaustive, in this
subsection we discuss the subtle points arising in the course of the
derivation of the residue formula when two or all the three differentials
$dv_I$ are associated
with punctures (masses). This is the case, when the $\Sigma$-term (\ref{sigma})
contributes to the result. In what follows we demonstrate
how this specific case of the residue formula can be treated in the example
of the theory with matter hypermultiplet in the fundamental representation,
and we start with consideration of the perturbative limit. To match the
notations of sect.5.2, we use the
generating differential $dS=\lambda{d\hat w\over\hat w}=-{1\over 2}\lambda
d\log{P-Y\over P+Y}$ and, accordingly, the holomorphic differentials
$dW_{\iota}^{(2)} \cong -{P\over 2Y}\frac{d\lambda}{(\lambda - m_{\iota})} -
\frac{\partial R}{\partial m_{\iota}}{d\lambda\over Y}$. Then,
in the perturbative limit
this $dW_{\iota}$ acquires the form (\ref{massdiff}),
$s_{\iota}=\lambda_{\iota}$.

First, we derive formula (\ref{resch})
in the perturbative limit with two mass
differentials. The symmetry of the l.h.s. of (\ref{resch})
w.r.t. the permutations of
indices is not obvious, since there are no symmetric terms in the l.h.s.
of (\ref{resch}) and one needs
to consider the entire sum. Therefore, in illustrative purposes, we
discuss here all possible arranging of indices in (\ref{resch}) --
$\{KLM\}=\{\iota,\iota,\iota'\}$, $\{KLM\}=\{\iota,\iota',\iota\}$ and
$\{KLM\}=\{\iota',\iota,\iota\}$ checking that they lead to the same result.
Note that the term $\oint_{A_I}{\partial dv_L\over\partial s_M}
\int_{{\d B_I\over \d s_K}}dS$ is zero in all the cases considered below.
Other terms are equal to
\be\label{sigma1}
\Sigma_{KLM}^{(1)}\equiv
\stackreb{\frac{\partial v_L}{\partial \lambda_M}=0}{\res}
\left(dv_K \frac{\partial v_L}{\partial \lambda_M}\right) =
\left\{\begin{array}{cl}
0     & \hbox{when}\ \ \ K=L    =\iota,\ M=\iota'
\\
0      &  \hbox{when}\ \ \ K=M  =\iota,\ L=\iota'
\\
-{1\over 4}{1\over\lambda_{\iota\iota'}} &  \hbox{when}\ \ \ L=M =\iota,\
K=\iota'
\end{array} \right.
\ee

\be
\Sigma_{KLM}^{(2)}\equiv\sum_I\oint_{A_I}dv_K {\p s_M}\int_{{\d
B_I\over \d \lambda_L}}dS=
\left\{\begin{array}{cl}
-{1\over 4}{1\over \lambda_{\iota\iota'}}& \hbox{when}\ \ \
K=L=\iota,\ M=\iota'
\\
0      &  \hbox{when}\ \ \  K=M=\iota,\ L=\iota'
\\
0      &  \hbox{when}\ \ \ L=M=\iota,\ K=\iota'
\end{array}
\right.
\ee

\be
\Sigma_{KLM}^{(3)}\equiv \sum_I
\oint_{A_I}dv_K \int_{{\d B_I\over \d \lambda_M}}dv_L=
\left\{\begin{array}{cl}
0     & \hbox{when}\ \ \ K=L    =\iota,\ M=\iota'
\\
-{1\over 4}{1\over \lambda_{\iota\iota'}}&  \hbox{when}\ \ \
K=M  =\iota,\ L=\iota'
\\
0       &  \hbox{when}\ \ \ L=M =\iota,\ K=\iota'
\end{array}
\right.
\ee
since $\oint_{A_I}dv_{\iota}=-{1\over 2}\delta_{I\iota}$,
$\int_{{\d B_{I}\over \d \lambda_{\iota}}}dv_{\iota'}=
{1\over 2}\delta_{I\iota}{1\over \lambda_{\iota\iota'}}$ and
$\int_{{\d B_I\over \d \lambda_{\iota}}}dS=
\left.\delta_{I\iota}\log\left({P(\lambda)\over
\sqrt{Q(\lambda)}}\right)\right|_{\lambda=
\lambda_{\iota}}$.

The sum of $\Sigma^{(1)}+\Sigma^{(2)}+\Sigma^{(3)}$
in each of the three cases $K=L$, $K=M$ and $L=M$
is the same how it is
required by the symmetricity w.r.t. the permutations of indices. This sum
is actually equal to $F_{\iota\iota\iota'}$ in (\ref{perth}).

Analogously, one can check the residue formula for three coinciding mass
differentials. Then, the result looks like
\be
\Sigma_{\iota\iota\iota}^{(1)}=-{1\over 4}\res {1\over (\lambda-
\lambda_{\iota})^2}=0
\\
\Sigma_{\iota\iota\iota}^{(2)}={1\over 2}\sum_i{1\over\lambda_{i\iota }}
+{1\over 4}\sum_{\iota'\ne\iota}{1\over\lambda_{\iota\iota'}}
+\left.{1\over 4}{1\over\lambda-\lambda_{\iota}}
\right|_{\lambda\to\lambda_{\iota}}
\\
\Sigma_{\iota\iota\iota}^{(3)}=-\left.
{1\over 4}{1\over\lambda-\lambda_{\iota}}
\right|_{\lambda\to\lambda_{\iota}}
\ee
and the sum of these terms is equal to $F_{\iota\iota\iota}$ in
(\ref{perth}).  Let us note that the only role of $\Sigma^{(3)}$ is to cancel
the singularity arising in $\Sigma^{(2)}$.

Now we can turn to less trivial case and discuss the residue formula in the
hyperelliptic case of sect.5.6. Again, we start with consideration of
$F_{\iota'\iota\iota}$. However, this time we restrict ourselves only to
the case $L=M$ -- the most convenient choice of order of the indices, since
it is the case when the only first term in the l.h.s. of (\ref{resch})
contributes. In the hyperelliptic case the residue acquires two different
contributions. The first one comes from the pole at $\lambda=m_{\iota}$
and can be
obtained with the help of expansion
\be\label{vsm}
{\d v_{\iota}\over\d m_{\iota}}\stackreb{\lambda
\to m_{\iota}}{\longrightarrow}{1\over 2}{1\over\lambda
-m_{\iota}}+{\cal O}
(\lambda- m_{\iota})
\ee
Therefore, the contribution of this pole is equal to that in
the perturbative case (\ref{sigma1}):
${1\over 4}{1\over \lambda_{\iota'\iota}}$. However, now there emerge also
some contributions from the ramification points. Indeed,
\be
{\d dv_{\iota}\over\d m_{\iota}} \stackreb{\lambda\to \lambda_{\alpha}}
{\longrightarrow} {1\over 2}\left({P(\lambda_{\alpha})\over
2(\lambda_{\alpha}-m_{\iota})}+{\d R(\lambda_{\alpha})
\over\d m_{\iota}}\right)
{\d Y^2\over\d m_{\iota}}{d\lambda\over Y^3}
={P(\lambda_{\alpha})\over
\hat Y^3(\lambda_{\alpha})}
\left({P(\lambda_{\alpha})\over
2(\lambda_{\alpha}-m_{\iota})}+{\d R(\lambda_{\alpha})
\over\d m_{\iota}}\right)
{d\lambda\over (\lambda-\lambda_{\alpha})^{3/2}}
\ee
since ${\d Y^2\over \d m_{\iota}}=2P{\d R\over\d m_{\iota}}+
{Q\over (\lambda-m_{\iota})}$. Therefore,
\be
{\d v_{\iota}\over\d m_{\iota}} \stackreb{\lambda\to \lambda_{\alpha}}
{\longrightarrow}
-2{P(\lambda_{\alpha})\over \hat Y^4(\lambda_{\alpha})}
\left({P(\lambda_{\alpha})\over
2(\lambda_{\alpha}-m_{\iota})}+{\d R(\lambda_{\alpha})
\over\d m_{\iota}}\right)^2
{1\over \sqrt{\lambda-\lambda_{\alpha}}}
\ee
and the full answer becomes
\be
F_{\iota\iota\iota'}=\Sigma_{\iota'\iota\iota}^{(1)}=2
\sum_{\alpha}
{P(\lambda_{\alpha})\over\hat Y^4(\lambda_{\alpha})}
\left({P(\lambda_{\alpha})\over
2(\lambda_{\alpha}-m_{\iota})}+{\d R(\lambda_{\alpha})
\over\d m_{\iota}}\right)^2
\left({P(\lambda_{\alpha})\over
2(\lambda_{\alpha}-m_{\iota'})}+{\d R(\lambda_{\alpha})
\over\d m_{\iota'}}\right)
-{1\over 4}{1\over \lambda_{\iota\iota'}}\label{fghb}
\ee
This result should be compared with the residue formula
\be
F_{\iota\iota\iota'}=\stackreb{{dw\over w}=0}{\res} {(dv_{\iota})^2
dv_{\iota'}\over d\lambda {dw\over w}}=\stackreb{{dw\over w}=0}{\res}
{1\over Y^2P}
\left({P\over
2(\lambda-m_{\iota})}+{\d R\over\d m_{\iota}}\right)^2
\left({P\over
2(\lambda-m_{\iota'})}+{\d R\over\d m_{\iota'}}\right)
{d\lambda\over
\left(\log(P/\sqrt{Q})\right)'}
\ee
As we did it earlier, instead of quite involved calculation of this residue
at zeroes of ${dw\over w}$, we calculate it in all other poles (i.e.
zeroes of $d\lambda$). Since $\left(\log(P/\sqrt{Q})\right)'$ contains
$\left[(\lambda-m_{\iota})(\lambda -m_{\iota'})\right]^{-1}$,
only ramification points and
$\lambda=m_{\iota}$ contributes to the result. Using formula
\be\label{aux1}
\left.\left(\log(P/\sqrt{Q})\right)'\right|_{\lambda=\lambda_{\alpha}}
={\hat Y^2(\lambda_{\alpha})\over
2P^2(\lambda_{\alpha})}
\ee
one can easily check that this really gives (\ref{fghb}).

Analogously one can
study in the hyperelliptic case the residue formula for the
three coinciding differentials $F_{\iota\iota\iota}$. This time the
calculation is a little longer. Using (\ref{vsm}), one gets for
$\Sigma^{(1)}_{\iota\iota\iota}$
\be\label{sigma1mmm}
\Sigma^{(1)}_{\iota\iota\iota}=
2\sum_{\alpha}
{P(\lambda_{\alpha})\over\hat Y^4(\lambda_{\alpha})}
\left({P(\lambda_{\alpha})\over
2(\lambda_{\alpha}-m_{\iota})}+{\d R(\lambda_{\alpha})
\over\d m_{\iota}}\right)^3
-{1\over 2P(m_{\iota})}{\d R(m_{\iota})\over
\d m_{\iota}}
-\left.{1\over 4}{\d(P/Y)\over \d\lambda}\right|_{\lambda=m_{\iota}}
\ee
i.e. the sum over ramification points looks similar
to the case of two coinciding differentials. Other $\Sigma$ terms are
even simpler:
\be\label{sigma2mmm}
\Sigma^{(2)}_{\iota\iota\iota}=
-\left.{1\over 2}{\d P/\d\lambda\over P}\right|_{\lambda=m_{\iota}}
-{1\over 2P(m_{\iota})}{\d R(m_{\iota})\over
\d m_{\iota}}
+{1\over 4}\sum_{\iota'\ne\iota}{1\over\lambda_{\iota\iota'}}
+\left.{1\over 4}{1\over\lambda-\lambda_{\iota}}
\right|_{\lambda\to\lambda_{\iota}}
\ee
\be\label{sigma3mmm}
\Sigma^{(3)}_{\iota\iota\iota}=-{1\over 2P(m_{\iota})}{\d R(m_{\iota})\over
\d m_{\iota}} -\left.
{1\over 4}{1\over\lambda-\lambda_{\iota}}
\right|_{\lambda\to\lambda_{\iota}}
\ee
The sum of all the three terms should be compared with the residue formula
\be\label{resmmm}
F_{\iota\iota\iota}=\stackreb{{dw\over w}=0}{\res} {(dv_{\iota})^3
\over d\lambda {dw\over w}}=-\stackreb{{dw\over w}=0}{\res}
{1\over Y^2P}
\left({P\over
2(\lambda-m_{\iota})}+{\d R\over\d m_{\iota}}\right)^3
{d\lambda\over
\left(\log(P/\sqrt{Q})\right)'}
\ee
Again taking residues at zeroes of $d\lambda$ and using formulas
(\ref{aux1}) along with
\be
\left.(\lambda-m_{\iota})\left(\log{P\over\sqrt{Q}}\right)'
\right|_{\lambda=m_{\iota}}=-{1\over 2}
\\
\left[\left.(\lambda-m_{\iota})\left(\log{P\over\sqrt{Q}}\right)'\right]'
\right|_{\lambda=m_{\iota}}=\left.{\d P/\d\lambda\over P}
\right|_{\lambda=m_{\iota}}-{1\over 2}
\sum_{\iota'\ne\iota}{1\over\lambda_{\iota\iota'}}
\ee
one finally proves that (\ref{resmmm}) is really equal to the sum of
(\ref{sigma1mmm}), (\ref{sigma2mmm}) and (\ref{sigma3mmm}).
Contributions from the
ramification points into (\ref{resmmm}) entirely originate
from analogous term in $\Sigma^{(1)}_{\iota\iota\iota}$, and
all the terms in $\Sigma$'s proportional to $\d R/\d m_{\iota}$
come from the first order pole at $\lambda=m_{\iota}$. The remaining
terms in $\Sigma^{(1)}$ and $\Sigma^{(2)}$ are equal to the residue of
the second order pole at $\lambda=m_{\iota}$.

\section{Hypermultiplet in the adjoint representation (Calogero
system) \label{Calogero}}
\setcounter{equation}{0}
\subsection{General formulas \label{genCa} }
To conclude our examples, in this section we discuss in detail the case of
Calogero system when
the closed algebra exists but is {\it not} associative, how it was
discussed in sect.3-4. The physics os this example is the
$4d$ ${\cal N}=2$ SUSY YM model with one matter hypermultiplet in the
adjoint representation of the gauge group $SU(N_c)$ that
is described in the SW
framework by elliptic Calogero model \cite{SWI3,IM1,IM2}. The bare
curve for the UV finite model is elliptic curve (\ref{torus}) and its modulus
$\tau $ can be identified with the bare (UV) coupling constant of the YM
theory. The mass of the $4d$ hypermultiplet is proportional to the
coupling constant $g$ in Calogero-Moser model. The spectral curve of
the elliptic Calogero model
\be
\det ({\cal L}(x,y) - \lambda ) = 0
\ee
is given by
\be\label{calocurve}
y^2 = (x-e_1(\tau))(x - e_2(\tau))(x-e_3(\tau)) =
x^3 - \frac{1}{4}g_2(\tau)x - \frac{1}{4}g_3(\tau) \nn \\
{\cal P}(\lambda ,x) = \sum _{i=0}^{N_c}g^is_i{\cal T}_i(\lambda, x) =0
\nn \\
{\cal T}_0 = 1, \ \ \ \ {\cal T}_1 = \lambda,\ \ \  \ \
{\cal T}_2 = \lambda^2 - x,
\ \ \ \ \ {\cal T}_3 = \lambda ^3 - 3\lambda x + 2y, \ \ \ \ \
{\cal T}_4 = \lambda^4 - 6\lambda^2 x + 8\lambda y - 3(x^2 + \mu_4),\nn \\
{\cal T}_5 = \lambda^5 - 10\lambda^3x + 20\lambda^2y -
15\lambda(x^2 + \mu_4) + 4xy, \ \ \ \ \
{\cal T}_6 = \lambda^6 - 15\lambda^4x + 40\lambda^3y
- 45\lambda^2(x^2 + \mu_4) + 24\lambda xy - 5,
\nn \\
\ldots
\label{explT}
\ee
(we remind that $e_1+e_2+e_3=0$).
Using gradation $\deg\ \lambda\ = 1$, $\deg\ x\ = 2$,
$\deg\ y\ =3$, $\ldots$,
${\cal T}_i$ becomes a homogeneous polynomial of degree $i$.
Actually this is a ``semi''-gradation from the point of view of
$x$ and $y$, because some coefficients also have non-vanishing
(but always positive) degree: $\deg\ e_i(\tau) = 2$,
$\deg\ \mu_4 = 4$ etc.
As functions of $\lambda$, the ${\cal T}_i$ behave similar to the
Schur polynomials:
\be\label{T'}
{\cal T}_i' \equiv \frac{\partial{\cal T}_i}{\partial \lambda} =
i{\cal T}_{i-1}.
\ee
As a corollary, ${\cal P}'$ is again a linear combination of
${\cal T}_i$:
\be
{\cal P}' \equiv \frac{\partial{\cal P}}{\partial\lambda} =
\sum_i ig^is_i{\cal T}_{i-1}.
\label{calP'}
\ee
For our purposes we can absorb $g_i$ in (\ref{calocurve}) into
the moduli $s_i$.

The full spectral curve (\ref{calocurve}) is a $N_c$-fold
covering of bare elliptic curve, which has genus $N_c$
(not $\sim N_c^2/2$ as one could naively expect), due to specific
properties of the functions ${\cal T}_i$ which may be demonstrated in
the Hitchin-like interpretation
of Calogero model, when the matrix-valued moment map is restricted
to have a pole with $N_c-1$ coinciding eigenvalues at it.
In the limit $x \sim \frac{1}{\xi^2} \rightarrow \infty$,
the coordinate $\lambda \sim \frac{1}{\xi}$ on the $N_c-1$ sheets
and $\lambda \sim -\frac{N_c-1}{\xi}$ on the last sheet. We shall
refer to these to kinds of sheets as ``$+$'' and ``$-$'' respectively.
On the $N_c-1$ ``$+$'' sheets all the functions ${\cal T}_i$
behave as
\be
{\cal T}_i \sim \lambda
\label{asympt}
\ee
in the limit $\lambda \longrightarrow \infty$, while on the ``$-$'' sheet
${\cal T}_i \sim \lambda^i$. Since ${\cal P}$ and ${\cal P'}$ are both
linear combinations of ${\cal T}_i$, they have the same asymptotics
$\sim\lambda^1$ on the ``$+$'' sheet.

For the generating form $dS=\lambda d\omega = \lambda
\frac{dx}{y}$, we have
\be
dv_k = \frac{\partial dS}{\partial s_k} =
-\frac{{\cal T}_k d\omega}{{\cal P}'}, \ \ \
k = 0,1,2,\ldots,N_c-2.
\label{calodiffs}
\ee
The complete basis of $N_c$ holomorphic 1-differentials on ${\cal C}$
consists of $dv_k$ with $k=0,1,2,\ldots,N_c-2,N_c-1$.
The holomorphic 1-differential
$dv_{N_c-1}$ is (up to addition of other $dv_k$ with $k\leq N_c-2$)
equal to $d\omega = \frac{dx}{y}$. The only non-vanishing
integrals of $d\omega$ are along the cycles on the {\it bare}
spectral curve $E(\tau)$. The periods $a_i$ relevant for the
SW theory are instead associated with the other cycles,
which turn to be
trivial after projection on $E(\tau)$ (see sect.8 of \cite{IM2} for discussion
of the $N_c=2$ example).

In other words, the polynomial ${\cal T}_{N_c-1}$ is linearly equivalent
to a combination of ${\cal T}_i$ with $i<N_c-1$ {\it modulo} ${\cal P}'$.
At the same time this is not true for ${\cal T}_{N_c}$, despite it is
of degree $N_c$ in $\lambda$, while ${\cal P}'$ is only of degree
$N_c-1$ -- this is because ${\cal T}_i$ are polynomials of several variables.
(In fact ${\cal T}_i$ is a pair of homogeneous polynomials of two variables,
$\lambda$ and $x$, it is a {\it pair}, because coefficients in front
of $y^0$ and $y^1$ are independent polynomials.)

One does not expect such polynomials to form a closed algebra
modulo a single polynomial ${\cal P}'$,
\be
{\cal T}_i{\cal T}_j = C_{ij}^k{\cal T}_k{\cal T}\
{\rm mod}\ {\cal P}'
\ee
instead one would rather expect a closed algebra modulo {\it two}
polynomials,
${\cal P}' = \partial{\cal P}/\partial\lambda$ and
$\dot{\cal P} = y\partial{\cal P}/\partial x$,
\be
{\cal T}_i{\cal T}_j = C_{ij}^k{\cal T}_k{\cal T}\
{\rm mod}\ ({\cal P}',\dot{\cal P}), \ \ \ \
{\cal T} = \sum_{k=0}^{N_c} q_k{\cal T}_k
\label{calal}
\ee
Such an algebra indeed exists, due to the general arguments
of sect.4. Technically this happens because of specific
properties  (\ref{T'}) and (\ref{asympt})  of
${\cal T}_k$.

The main thing is that among quadratic combinations
${\cal T}_i{\cal T}_j$ there are only few linear
independent. Because of the (semi) gradation ${\cal T}_i{\cal T}_j$
can be expressed only through the combinations of the same
(or smaller) degree $r = i+j$, and actually one can express
everything in terms of, say, ${\cal T}_0{\cal T}_r =
{\cal T}_r$ and ${\cal T}_1{\cal T}_{r-1} = \lambda{\cal T}_{r-1}$.
This of course follows from the counting of holomorphic
quadratic differentials, but can be also checked directly,
using explicit expressions (\ref{explT}) for ${\cal T}_k$.
The first linear independent combinations are:
\be
\begin{array}{lll}
{\rm deg}\ 0:& {\cal T}_0 = 1;& \\
{\rm deg}\ 1:& {\cal T}_1 = \lambda; & \\
{\rm deg}\ 2:& {\cal T}_2 = \lambda^2-x,\ {\cal T}_1^2 = \lambda^2; & \\
{\rm deg}\ 3:& {\cal T}_3, \ {\cal T}_1{\cal T}_2;& \\
{\rm deg}\ 4:& {\cal T}_4, \ {\cal T}_1{\cal T}_3,\ {\cal T}_2^2;&
  {\cal T}_4 = {\cal T}_1{\cal T}_3 - 3{\cal T}_2^2 - 3\mu_4; \\
{\rm deg}\ 5:& {\cal T}_5, \ {\cal T}_1{\cal T}_4,\ {\cal T}_2{\cal T}_3; &
  {\cal T}_5 = 3{\cal T}_1{\cal T}_4 - 2{\cal T}_2{\cal T}_3
- 6\mu_4{\cal T}_1;
\end{array} \\
\ldots
\ee

As we shall see in the subsequent subsections, the combination
$\tilde{\cal P}_\omega \equiv \dot{\cal P} - x{\cal P}'$
is always a linear combination of ${\cal T}_i{\cal T}_j$,
and this is enough to make the algebra (\ref{calal}) {\it closed}.
We demonstrate this explicitly for the first non-trivial cases
of $N_c=3$ and $N_c=4$. However, the algebra is non-associative,
starting from $N_c=4$, and again this is checked by explicit
calculation (which is somewhat tedious and is actually performed
with the help of computer). These examples can also help to
illustrate some (possibly obscure) aspects of the reasoning
in sect.4. Moreover, we begin from the case $N_c=2$, which is also
instructive for the future studies of the {\it physical} properties
of the UV-finite theories. Results in this case can be
directly compared with those of instanton calculus -- once
they will be obtained for the case of matter in the adjoint
representation. Of course, nothing interesting can be
said about the WDVV equations and even the algebra of differentials
for $N_c=2$, but one can shed new light on residue formulas --
as well as on some other aspects of the SW theory.

\subsection{Residue formula: example for  $N_c=2$}

In the case of the gauge group $SU(2)$
the full spectral curve is the double covering of the torus $E(\tau)$:
\be
{\cal P}^{N_c = 2}(\lambda ;x,y) = \lambda^2 - g^2 x + s_0 = 0
\ee
The generating differential is
\be
dS = \lambda d\omega = \frac{\sqrt{s_0 - g^2 x} dx}{y}
\ee
and the prepotential is given by
\be
F = \frac{1}{2}\left( \oint_A dS \oint_B dS +
\oint_{\lambda = \infty} dS
\left.\left(
\int_{\lambda = -\infty+\varepsilon }^{\lambda = +\infty - \varepsilon} dS
- g\log\varepsilon\right)\right|_{\varepsilon = 0}\right)
\label{prepN2}
\ee
The derivative
\be
dv =
\frac{\partial dS}{\partial s_0} = \frac{d\omega}{2\lambda} =
\frac{dx}{2\lambda y} = \frac{dx}{2y\sqrt{s_0 - g^2 x}}
\ee
and
\be
\frac{\partial a}{\partial s_0} = \frac{\partial}{\partial s_0}
\oint_A dS = \oint_A \frac{dx}{2\lambda y} \ \ \ \ \ \
\frac{\partial}{\partial a} =
\left(\oint_A \frac{dx}{2\lambda y}\right)^{-1}
\frac{\partial}{\partial s_0}
\ee
Since $dS$ and $dv$ are both $(1,0)$-differentials,
$dS\wedge dv = 0$, and
\be
0 = \int_{{\cal C}}dS\wedge dv = \oint_A dS \oint_B dv  - \oint_B dS\oint_A dv
+ \oint_{\lambda = \infty} dS
\int_{\lambda = -\infty }^{\lambda = +\infty} dv
\label{prome1}
\ee
The r.h.s. is the result of application of the Stokes theorem to
the total derivative $dS\wedge dv = d(Sdv)$ on a simply-connected
surface, obtained by cutting the spectral curve ${\cal C}$
along $A$- and $B$-cycles and, additionally, between the two
points $\lambda = \pm\infty$, where $dS$ (but not $dv$) has
simple poles. Eq.(\ref{prome1}) states that
\be
a\frac{\partial a^D}{\partial s_0} - a^D\frac{\partial a}
{\partial s_0} +  g\int_{-\infty}^{+\infty} dv = 0,
\ee
thus
\be
\frac{\partial F}{\partial s_0} = \frac{1}{2}\left(
a\frac{\partial a^D}{\partial s_0} + a^D\frac{\partial a}
{\partial s_0} +  g\int_{-\infty}^{+\infty} dv  \right) =
a^D\frac{\partial a}{\partial s_0},
\ee
i.e.
\be
\frac{\partial F}{\partial a} = a^D
\ee
and
\be
\frac{\partial^2 F}{\partial a^2} = \frac{\partial a^D/\partial s_0}
{\partial a/\partial s_0}
= \frac{\oint_B \frac{dx}{\lambda y}}{\oint_A \frac{dx}{\lambda y}}
\ee
so that
\be
\frac{\partial^3 F}{\partial a^3} =
\left(\oint_A \frac{dx}{\lambda y} \right)^{-3}
\left(\oint_A \frac{dx}{\lambda y} \oint_B \frac{dx}{\lambda^3 y}
- \oint_B \frac{dx}{\lambda y} \oint_A \frac{dx}{\lambda^3 y}\right)
\label{promm1}
\ee
The next is to express the difference in brackets through the residue at
$\lambda = 0$:
\be
\frac{\partial^3 F}{\partial a^3} =
2\pi i\left(\oint_A \frac{dx}{\lambda y} \right)^{-3}
\stackreb{\lambda = 0}{\res} \frac{dx}{\lambda^3 y} d^{-1}
\frac{dx}{\lambda y} =
\left(\oint_A \frac{dx}{2\lambda y} \right)^{-3}
\frac{2\pi i}{2y^2\left(x=\frac{s_0}{g^2}\right)}
\label{prome2}
\ee
(we used the fact that on ${\cal C}$ the differential
$dx = \frac{2\lambda d\lambda}{g^2}$, and there is an extra factor
$2$ coming from two sheets of $E(\tau)$).

For illustrative purposes let us represent all formulas
in a different way: expand in powers of $g^2$ using
\be
\lambda = \sqrt{s_0-g^2 x} = s_0^{1/2}\sum_{k=0}^\infty
\frac{\Gamma(3/2)}{k!\Gamma(3/2-k)} \left(-\frac{g^2 x}{s_0}\right)^k
= s_0^{1/2}\left(1 - \sum_{k=1}^\infty \frac{(2k-3)!!}{k!}
\left(\frac{g^2x}{2s_0}\right)^k\right)
\nn \\
\frac{1}{\lambda} = \frac{1}{\sqrt{s_0-g^2 x}} =
s_0^{-1/2}\sum_{k=0}^\infty
\frac{\Gamma(1/2)}{k!\Gamma(1/2-k)} \left(-\frac{g^2 x}{s_0}\right)^k
= s_0^{-1/2} \sum_{k=0}^\infty \frac{(2k-1)!!}{k!}
\left(\frac{g^2x}{2s_0}\right)^k
\nn \\
\frac{1}{\lambda^3} = \frac{1}{(s_0-g^2 x)^{3/2}} =
s_0^{-3/2}\sum_{k=0}^\infty
\frac{\Gamma(-1/2)}{k!\Gamma(-1/2-k)} \left(-\frac{g^2 x}{s_0}\right)^k
= s_0^{-3/2} \sum_{k=0}^\infty \frac{(2k+1)!!}{k!}
\left(\frac{g^2x}{2s_0}\right)^k
\ee
($(-1)!! = 0! = 1$).
Among contour integrals $\oint \frac{x^kdx}{y}$ (for a given homotopy
class of contours) there are only two independent: for example with $k=0$ and
$k=1$. All others can be expressed through these ones, using relations
\be
dy = \frac{dx}{2y}(3x^2 - \frac{g_2}{4})
\nn \\
d(x^{k-2}y) = \frac{dx}{y}\left( \frac{3}{2}x^k - \frac{g_2}{8}x^{k-2}
+ (k-2)x^k - (k-2)\frac{g_2}{4}x^{k-2} - (k-2)\frac{g_3}{4}x^{k-3}\right)
=  \nn \\ =
(k-\frac{1}{2})\frac{dx}{y} \left( x^k - \frac{2k-3}{4(2k-1)}g_2x^{k-2}
- \frac{k-2}{2(2k-1)}g_3 x^{k-3}\right)
\ee
Thus, one can substitute:
\be
\oint dv = \rho \oint\frac{dx}{2y} + \sigma \oint\frac{xdx}{2y},\nn \\
\rho = \frac{1}{s_0^{1/2}}\left( 1 +
\frac{g_2}{8}\left(\frac{g^2}{2s_0}\right)^2 +
\frac{g_3}{4}\left(\frac{g^2}{2s_0}\right)^3 +
\frac{25g_2^2}{384}\left(\frac{g^2}{2s_0}\right)^4 +
\frac{21g_2g_3}{80}\left(\frac{g^2}{2s_0}\right)^5 + \ldots\right)
\nn \\
\sigma = \frac{1}{s_0^{1/2}}\left(
\left(\frac{g^2}{2s_0}\right) +
\frac{3g_2}{8}\left(\frac{g^2}{2s_0}\right)^3 +
\frac{5g_3}{8}\left(\frac{g^2}{2s_0}\right)^4 +
\frac{147g_2^2}{640}\left(\frac{g^2}{2s_0}\right)^5 +
\ldots\right)
\label{prome3}
\ee
The conventional notation for the remaining periods in the theory of
elliptic functions is:
\be
\oint_A \frac{dx}{2y} = \oint_A d\xi = 2\omega \ \ \
\oint_B \frac{dx}{2y} = 2\omega' = 2\omega\tau \nn \\
\oint_A \frac{xdx}{2y} = \oint_A \wp(\xi)d\xi =
\eta(\xi + 2\omega) - \zeta(\xi) = 2\hat\eta \equiv 2\omega\bar\eta \ \ \
\oint_B \frac{xdx}{2y} = 2\hat\eta' \equiv 2\omega\bar\eta' \nn \\
\omega' = \omega\tau \ \ \ \
\bar\eta\omega' - \bar\eta'\omega =
\omega^{-1}(\hat\eta\omega' - \hat\eta'\omega) =
\frac{i\pi}{2\omega}
\ee
so that
\be
\frac{\partial^2 F}{\partial a^2} = \frac{\oint_B dv}{\oint_A dv} =
\frac{\rho\omega' + \sigma\hat\eta'}{\rho\omega + \sigma\hat\eta} =
\tau -\frac{i\pi}{2\omega^2} \frac{\sigma}{\rho + \sigma\bar\eta}
\label{chargeN2}
\ee
and
\be
\frac{\partial^3 F}{\partial a^3} =
-\frac{i\pi}{2\omega^2}(\rho + \sigma\bar\eta)^{-2}
\left(\rho\frac{\partial \sigma}{\partial a} - \sigma
\frac{\partial \rho}{\partial a}\right) =
-\frac{i\pi}{4\omega^3}(\rho + \sigma\bar\eta)^{-3}
\left(\rho\frac{\partial \sigma}{\partial s_0} - \sigma
\frac{\partial \rho}{\partial s_0}\right)  = \nn \\ =
2\pi i \left(\oint_A dv\right)^{-3}
\left(\rho\frac{\sigma{\partial \rho}
- \rho \partial \sigma}{\partial s_0}  \right)
\ee
In order to reproduce (\ref{prome2}), it remains to check that
the difference in brackets is indeed equal to
\be
\frac{1}{2g^4y^2(x = \frac{s_0}{g^2})} =
\frac{g^2}{2\prod_{i=1}^3 (s_0 - g^2 e_i(\tau))} =
\nn \\
= \frac{g^2}{2s_0^3}\left(1 +  g_2\left(\frac{g^2}{2s_0}\right)^2 +
2g_3 \left(\frac{g^2}{2s_0}\right)^3 +
g_2^2\left(\frac{g^2}{2s_0}\right)^4 +
4g_2g_3\left(\frac{g^2}{2s_0}\right)^5 + \ldots\right)
\ee
what is easy to do with the help of expansions (\ref{prome3}).

From (\ref{chargeN2}), with
\footnote{The origin of the shift
$s_0 \rightarrow \tilde s_0$
is explained in detail in \cite{IM1,IM2}.}
\be
\alpha \equiv -\frac{m^2}{16s_0} =
\left(\frac{\pi}{2\omega}\right)^2\frac{g^2}{2s_0} =
\frac{g^2}{2(g^2 \bar\eta + \tilde s_0)} =
\frac{\tilde\alpha}{1 + 2\eta\tilde\alpha}
\ee
$\tilde\alpha \equiv -\frac{m^2}{16\tilde s_0}$,
and
\be
\eta \equiv \left(\frac{2\omega}{\pi}\right)^2\frac{\hat\eta}{\omega}
\nn \\
\frac{\hat\eta}{\omega} = - \frac{\pi^2}{12 w_1^2} \frac{\theta_{11}'''(0)}
{\theta_{11}'(0)} =
\left(\frac{\pi}{2\omega}\right)^2\left(\frac{1}{3} + O(q^2)\right) \nn \\
g_2 = \left(\frac{\pi}{2\omega}\right)^4\left(\frac{4}{3} + O(q^2)\right)
\nn \\
g_3 = \left(\frac{\pi}{2\omega}\right)^6\left(-\frac{8}{27} + O(q^2)\right)
\nn \\
\ldots
\ee
we get:
\be
T \equiv \frac{\partial^2F}{\partial a^2} =
\tau - \frac{2i}{\pi}\left(\alpha - \eta\alpha^2 +
(\eta^2 +\frac{g_2}{4})\alpha^3 -
(\eta^3 + \frac{\eta g_2}{2} - \frac{3g_3}{8})\alpha^4
+ (\eta^4 + \frac{3\eta^2g_2}{4} - \frac{3\eta g_3}{4}
+ \frac{2g_2^2}{15}) + \ldots \right) = \nn \\ =
\tau - \frac{2i}{\pi} \left( \tilde\alpha - 3\eta\tilde\alpha^2 +
(9\eta^2 + \frac{g_2}{4})\tilde\alpha^3 -
(27\eta^3 + 2\eta g_2 - \frac{3g_3}{8})\tilde\alpha^4 +
(81\eta^4 + \frac{43}{4}\eta^2 g_2 - \frac{15}{4}\eta g_3 +
\frac{2}{15}g_2^2)\tilde\alpha^5 + \ldots\right) = \nn \\ =
\tau - \frac{2i}{\pi} \left(\tilde\alpha - \tilde\alpha^2 +
\frac{4}{3}\tilde\alpha^3 - 2\tilde\alpha^4 + \frac{16}{5}
\tilde\alpha^5 + \ldots\right) + O(q^2) = \nn \\ =
\tau - \frac{i}{2\pi}\log (1 + 2\tilde\alpha) + O(q^2)
\label{expan}
\ee
Further,
\be
a = \oint_A\lambda\frac{dx}{2y}  = s_0^{1/2}\left(
1 - \frac{1}{24}\alpha^2g_2 - \frac{1}{20}\alpha^3g_3
- \frac{25}{12\cdot 8\cdot 28}\alpha^4g_2^2 - \ldots\right)
\oint_A\frac{dx}{2y} - \nn \\ -
s_0^{1/2}\left(\alpha + \frac{3}{40}\alpha^3g_2 +
\frac{5}{7\cdot 8}\alpha^4g_3 + \ldots\right)\oint_A
\frac{xdx}{2y} =
\nn \\ =
s_0^{1/2}\left(1 - \alpha\eta - \frac{1}{24}\alpha^2g_2
- \alpha^3 (\frac{1}{20}g_3 + \frac{3}{40}\eta g_2) -
\alpha^4(\frac{25}{12\cdot 8\cdot 28}g_2^2 + \frac{5}{7\cdot 8}
g_3\eta) - \ldots\right) = \nn \\ =
s_0^{1/2}\sqrt{1-2\alpha\eta} + O(q^2) =
\tilde s_0^{1/2} + O(q^2)
\ee
Thus
\be\label{explog}
T = \tau - \frac{i}{\pi}\log\left(1 - \frac{m^2}{8a^2}\right)
+ O(q^2) = \tau + \frac{i}{\pi}\log\frac{a^2}{a^2 - \frac{1}{8}m^2} +
O(q^2)
\ee
This establishes the relation between residue formulas, the
prepotential and its perturbative approximation -- the logarithmic term in
 (\ref{explog}). It would be
instructive to find a careful interpretation of entire
expansion (\ref{expan}) in terms of instanton calculus in the
spirit of \cite{inst}.

\subsection{Example: Algebra for $N_c=3$}

In this case, there are only {\it two} relevant holomorphic
differentials: $dv_0$ and $dv_1$; the third one, $dv_2$, is
their linear combination {\it modulo} $d\omega = d\xi$.
Thus, there are no interesting WDVV equations for $N_c=3$,
but non-trivial associative algebra with two generators
$dv_0$ and $dv_1$ does exist and will be explicitly constructed
in this subsection.

In this particular case, the spectral curve is:
\be
{\cal P}^3 = {\cal T}_3 + s_1{\cal T}_1 + s_0{\cal T}_0 =
\lambda^3 - 3\lambda x + 2y + s_1\lambda + s_0 = 0
\ee
Thus
\be
{\cal P}^3_\lambda \equiv \frac{\partial{\cal P}_3}{\partial\lambda}
= 3{\cal T}_2 + s_1{\cal T}_0 =
3\left(\lambda^2 - x +\frac{s_1}{3}\right)
\ee
and
\be
{\cal P}^3_\omega \equiv \frac{\partial{\cal P}^3}{\partial\xi} =
-6\left(x^2 - \lambda y -\frac{g_2}{12}\right)
\nn \\
\tilde{\cal P}^3_\omega \equiv {\cal P}^3_\omega - x{\cal P}^3_\lambda
= -3\left(x^2 - 2\lambda y + \lambda^2 x + \frac{s_1}{3}x - \frac{g_2}{6}
\right) =
{\cal T}_4 - {\cal T}_1{\cal T}_3 - s_1{\cal T}_1 +
\left(3\mu_4 + \frac{g_2}{2}\right)
\ee
As $\xi \rightarrow 0$,  $x = \frac{1}{\xi^2} + O(\xi^2)$,
$y = \frac{1}{\xi^3} + O(\xi)$, and
\be
{\cal P}^3 = \left(\lambda - \frac{1}{\xi}\right)^2\left(\lambda
+ \frac{2}{\xi}\right) + s_1\lambda + O(\lambda\xi^2,\xi)
\ee
i.e.
\be
\lambda = \lambda_+ = \frac{1}{\xi} + \epsilon_0 + \epsilon_1\xi +
\epsilon_2\xi^2 + \ldots
\ee
with $2\epsilon_0^2 + s_1 = 0$ etc, or
\be
\lambda = \lambda_- = -\frac{2}{\xi} + \varepsilon_1\xi +
\varepsilon_3\xi^3 + \ldots
\ee
(Note that in the particular case of $N_c=3$ the point $\xi = 0$
is not a branching point of the full spectral curve and $\xi$ is
a nice local coordinate on all the three sheets.)

On the ``$+$'' sheets
\be
{\cal P}^3_\lambda = 3\left(\frac{2\epsilon_0}{\xi} +
(2\epsilon_1 + \epsilon_0^2 + \frac{s_1}{3}) +
(2\epsilon_2 + 2\epsilon_0\epsilon_1)\xi + O(\xi^2)\right) \nn \\
{\cal P}^3_\omega = 6\left(\frac{\epsilon_0}{\xi^3} +
\frac{\epsilon_1}{\xi^2} + \frac{\epsilon_2}{\xi} + O(1)\right)
\ee
and
\be
\tilde{\cal P}^3_\omega = {\cal P}^3_\omega - x{\cal P}^3_\lambda =
-\frac{3\epsilon_0^2 + s_1}{\xi^2} - \frac{6\epsilon_0\epsilon_1}{\xi}
+ O(1) = -\frac{6\epsilon_0\epsilon_1}{\xi} + O(1)
\ee
Thus
\be
d\lambda + xd\omega =
-\frac{\tilde{\cal P}^3_\omega}{{\cal P}^3_\lambda}d\omega
= (\epsilon_1 + O(\xi))d\xi
\ee
as it should be ($d\lambda + xd\omega = d(\frac{1}{\xi} + \epsilon_0
+ \epsilon_1\xi + \ldots) + (\frac{1}{\xi^2} + O(\xi^2))d\xi =
(\epsilon_1 + O(1))d\xi$).
On the ``$-$'' sheet
\be
{\cal P}^3_\lambda = \frac{9}{\xi^2} + O(1) \nn \\
{\cal P}^3_\omega = -\frac{18}{\xi^4} + O(\frac{1}{\xi^2})\nn \\
\tilde{\cal P}^3_\omega = {\cal P}^3_\omega - x{\cal P}^3_\lambda =
-\frac{27}{\xi^4} + O(\frac{1}{\xi^2})
\ee
again in accordance with
\be
d\lambda + xd\omega = -\frac{\tilde{\cal P}^3_\omega}{{\cal P}^3_\lambda}
d\omega = \left(\frac{3}{\xi^2} + O(1)\right)d\xi
\ee

\paragraph{\bf Algebra of holomorphic differentials}

\be
dv_i = \frac{{\cal T}_id\omega}{{\cal P}^3_\lambda}, \ \ \
i = 0,1,2
\ee
We claim that for any
\be
dQ = \hat q_0dv_0 + \hat q_1dv_1
\ee
and $i,j = 0,1$
\be
dv_idv_j = \left(\sum_{k=0,1} \hat C_{ij}^k dv_k\right) dQ
\ {\rm mod}\ (d\omega, d\lambda)
\label{algN3}
\ee
The notations with hats are introduced to point out that we consider the
algebra of $dv_i$'s, which are not {\it canonically} normalized holomorphic
differentials $d\omega_i$, but certain moduli-dependent linear combinations
of them. The fact that algebra exists for $dv_i$'s implies that the same is
true for $d\omega_i$'s, but explicit expressions for $dv_i$'s are simpler
(of course, $d\omega_i$'s themselves enter the residue formulas for
$F_{ijk}$).

At the l.h.s. of (\ref{algN3}) stands some linear combination of
{\it three} linear independent quadratic combinations
${\cal T}_0^2$, ${\cal T}_0{\cal T}_1$ and ${\cal T}_1^2$, i.e.
$1$, $\lambda$ and $\lambda^2$. At the r.h.s. we have:
\be
\left.
(\alpha_0{\cal T}_0 + \alpha_1{\cal T}_1)
(\hat q_0{\cal T}_0 + \hat q_1{\cal T}_1) +
w\tilde P^3_\omega \ {\rm mod}\ P^3_\lambda
\ \right|_{{\cal P}^3 = 0}
\ee
If $w=0$, there is no chance to adjust {\it two} parameters
$\alpha_0$ and $\alpha_1$ to match {\it three} coefficients in
front of ${\cal T}_0^2$, ${\cal T}_0{\cal T}_1$ and ${\cal T}_1^2$.
$w$ is exactly the necessary third adjustment parameter.
However, in order to prove that it can really play this role,
one should check that ${\cal P}^3_\omega = \tilde{\cal P}^3_\omega
\ {\rm mod}\ {\cal P}^3_\lambda$ is indeed equivalent to a non-trivial
combination of ${\cal T}_0^2$, ${\cal T}_0{\cal T}_1$ and ${\cal T}_1^2$
(if ${\cal P}^3 = 0$ and {\it modulo} ${\cal P}^3_\lambda$).

First of all,
\be
\left. \tilde {\cal P}^3_\omega \right|_{{\cal P}^3 = 0} =
3\left(\lambda^4 - 2\lambda^2 x + x^2 + s_1(\lambda^2 + \frac{x}{3})
+ s_0\lambda - \frac{g_2}{6}\right) = \nn \\
= 3{\cal T}_2^2 - s_1{\cal T}_2 + 4s_1{\cal T}_1^2 + 3s_0{\cal T}_1
- \frac{g_2}{2}
\ee
Alternatively, one can use
\be
\tilde{\cal P}^3_\omega = -\lambda {\cal P}^3 +
{\cal T}_4 + s_1{\cal T}_2 + s_0{\cal T}_1 +
3(\mu_4 + \frac{g_2}{2})
\ee
and
\be
{\cal T}_4 = {\cal T}_3{\cal T}_1 - 3{\cal T}_2^2 - 3\mu_4
\ee
Second,
\be
{\cal T}_2 = \frac{1}{3}({\cal P}^3_\lambda - s_1{\cal T}_0) =
- \frac{s_1}{3} \ {\rm mod}\ {\cal P}^3_\lambda, \nn \\
{\cal T}_2^2 = \frac{1}{9} ({\cal P}^3_\lambda - s_1{\cal T}_0)^2 =
\frac{s_1^2}{9} \ {\rm mod}\ {\cal P}^3_\lambda,
\ee
thus, indeed,
\be
\left. \tilde {\cal P}^3_\omega \ {\rm mod}\ {\cal P}^3_\lambda
\right|_{{\cal P}^3 = 0} =
\left(\frac{2s_1^2}{3} - \frac{g_2}{2}\right){\cal T}_0^2 +
3s_0 {\cal T}_0{\cal T}_1 + 4s_1{\cal T}_1^2,
\ee
and the r.h.s. of (\ref{algN3}) is equal to
\be
{\cal T}_0^2 \left(\hat q_0\alpha_0 +
           (\frac{2s_1^2}{3} - \frac{g_2}{2})w\right) + \nn \\
+{\cal T}_0{\cal T}_1 \left(\hat q_1\alpha_0 + \hat q_0\alpha_1
                     + 3s_0w\right) + \nn \\
+ {\cal T}_1^2 \left( \hat q_1\alpha_1 + 4s_1w\right)
\ee

For any given $dv_idv_j$ with $i,j=0,1$ at the l.h.s. of
(\ref{algN3}), one can now find $\alpha_0$, $\alpha_1$ and $w$
(since determinant $D = 4\hat q_0^2 s_1 - 3\hat q_0\hat q_1 s_0
+ \hat q_1^2 (\frac{2s_1^2}{3} - \frac{g_2}{2}) $ is non-vanishing
for generic values of moduli $s_0$ and $s_1$),
$\alpha_0$ and $\alpha_1$ being just the corresponding
$\hat C_{ij}^0$ and $\hat C_{ij}^1$. Namely, if the l.h.s.
of (\ref{algN3}) is given by
\be
a_{00}{\cal T}_0^2 + a_{01}{\cal T}_0{\cal T}_1 +
a_{11}{\cal T}_1^2 = a_{00} + a_{01}\lambda + a_{11}\lambda^2
\ee
we have:
\be
\left( \begin{array}{c} \alpha_0 \\ \alpha_1 \\ w \end{array}\right)
= \frac{1}{D} \left(\begin{array}{ccc}
4\hat q_0s_1 - 3\hat q_1s_0 &
             \hat q_1(\frac{2s_1^2}{3} - \frac{g_2}{2}) &
            -\hat q_0(\frac{2s_1^2}{3} - \frac{g_2}{2}) \\
-4\hat q_1s_1 & 4\hat q_0 s_1 &
    -3\hat q_0s_0 + \hat q_1(\frac{2s_1^2}{3} - \frac{g_2}{2}) \\
\hat q_1^2 & -\hat q_0\hat q_1 & \hat q_0^2 \end{array}\right)
\left(\begin{array}{c} a_{00}\\a_{01}\\a_{11}\end{array}\right)
\ee
In this way we obtain for the matrices $(\hat C_0)^k_j \equiv
\hat C_{0j}^k$ and $(\hat C_1)^k_j \equiv \hat C_{1j}^k$:
\be
\hat C_0 = \left(\begin{array}{cc} \alpha_0(1) & \alpha_0(\lambda) \\
            \alpha_1(1) & \alpha_1(\lambda) \end{array}\right) =
\frac{1}{D} \left(\begin{array}{cc}
   4\hat q_0s_1 - 3\hat q_1s_0 &
             \hat q_1(\frac{2s_1^2}{3} - \frac{g_2}{4}) \\
-4\hat q_1s_1 & 4\hat q_0 s_1  \end{array}\right), \nn \\
\hat C_1 = \left(\begin{array}{cc} \alpha_0(\lambda) & \alpha_0(\lambda^2) \\
            \alpha_1(\lambda) & \alpha_1(\lambda^2) \end{array}\right) =
\frac{1}{D} \left(\begin{array}{cc}
  \hat q_1(\frac{2s_1^2}{3} - \frac{g_2}{4}) &
            -\hat q_0(\frac{2s_1^2}{3} - \frac{g_2}{4}) \\
  4\hat q_0 s_1 &
             -3\hat q_0s_0 + \hat q_1(\frac{2s_1^2}{3} - \frac{g_2}{4})
\end{array}\right)
\ee
It is easy to check that these two matrices commute,
\be
\hat C_0 \hat C_1 = \hat C_1 \hat C_0
\ee
(in fact, $\hat q_0\hat C_0 + \hat q_1\hat C_1 = I$),
what is equivalent to associativity of the algebra
\be
dv_i \circ dv_j = \sum_{k=0,1} \hat C_{ij}^k dv_k
\ee

\subsection{Example: Algebra for  $N_c=4$}

In this case there are  {\it three} relevant holomorphic
differentials: $dv_0$, $dv_1$ and $dv_2$, while the fourth one, $dv_3$, is
their linear combination {\it modulo} $d\omega = d\xi$.
Thus, this is the simplest example where non-trivial WDVV equation could
exist. The purpose of this subsection is to give explicit realization of the
closed algebra with three generators $dv_0$, $dv_1$ and $dv_2$ which is
really {\it not} associative. We shall closely follow the presentation in
the previous subsection.

The spectral curve:
\be
{\cal P}^4 = {\cal T}_4 + s_2{\cal T}_2 + s_1{\cal T}_1 + s_0{\cal T}_0 =
\nn \\ =
\lambda^4 - 6\lambda^2 x + 8\lambda y - 3(x^2 + \mu_4)
+ s_2(\lambda^2-x) + s_1\lambda + s_0 = 0.
\ee
The choice of the $\tau$-dependent
parameter $\mu_4$ depends on the particular way of separating
$s_i$ and $\tau$ dependencies \cite{IM1}, which we keep unspecified.
\be
{\cal P}^4_\lambda \equiv \frac{\partial{\cal P}^4}{\partial\lambda}
= 4{\cal T}_3 + 2s_2{\cal T}_1 + s_1{\cal T}_0 = \nn \\ =
4\left(\lambda^3 - 3\lambda x + 2y +
\frac{s_2}{2}\lambda + \frac{s_1}{4}\right)
\ee
and
\be
{\cal P}^4_\omega \equiv \frac{\partial{\cal P}^4}{\partial\xi} =
12 \left(\lambda^2 y - 2\lambda (x^2 - \frac{g_2}{12}) + xy +
\frac{s_2}{6}y\right), \nn \\
\tilde{\cal P}^4_\omega \equiv {\cal P}^4_\omega - x{\cal P}^4_\lambda
= 4\left(xy - 3\lambda x^2 + 3\lambda^2 y - \lambda^3 x +
\frac{s_2}{2}(y-\lambda x) - \frac{s_1}{4}x + \frac{g_2}{2}\lambda
\right) = \nn \\ =
{\cal T}_5 - {\cal T}_1{\cal T}_4 + s_2({\cal T}_3 - {\cal T}_1{\cal T}_2)
+ 2(\mu_4 + g_2){\cal T}_1
\ee

\paragraph{The $\xi \rightarrow 0$ limit}

As $\xi \rightarrow 0$,  $x = \frac{1}{\xi^2} + O(\xi^2)$,
$y = \frac{1}{\xi^3} + O(\xi)$, and
\be
{\cal P}^4 = \left(\lambda - \frac{1}{\xi}\right)^3\left(\lambda
+ \frac{3}{\xi}\right) + s_2
\left(\lambda - \frac{1}{\xi}\right)\left(\lambda
+ \frac{1}{\xi}\right) + s_1\lambda + O(\lambda\xi),
\ee
i.e.
\be
\lambda = \lambda_+ = \frac{1}{\xi} + \epsilon_0 + \epsilon_1\xi +
\epsilon_2\xi^2 + \ldots;
\ee
with $4\epsilon_0^3 + 2s_2\epsilon_0^2 + s_1 = 0$ etc, {\it or}
\be
\lambda = \lambda_- = -\frac{3}{\xi} + \varepsilon_1\xi +
\varepsilon_2\xi^2 + \ldots
\ee
(Note that if all the moduli $s_i=0$, on the ``$+$'' branches
expansion would be rather in powers of $\xi^{1/3}$,
i.e. $\xi = 0$ would be a branching point where three of the four
sheets are glued together. We, however, assume that $s_i \neq 0$.)

On the ``$+$'' sheets
\be
{\cal P}^4_\lambda = 4\left(\frac{3\epsilon_0^2 + \frac{s_0}{2}}{\xi} +
(6\epsilon_0\epsilon_1 + \epsilon_0^3 + \frac{s_2}{2}\epsilon_0 +
\frac{s_1}{4}) +
(6\epsilon_0\epsilon_2 +  3\epsilon_1^2 +
\epsilon_0^2\epsilon_1 + \frac{s_2}{2}\epsilon_1)\xi +
O(\xi^2)\right); \nn \\
{\cal P}^3_\omega = 12\left(\frac{\epsilon_0^2 + \frac{s_2}{6}}{\xi^3} +
\frac{2\epsilon_0\epsilon_1}{\xi^2} + O(\frac{1}{\xi})\right);
\ee
and
\be
\tilde{\cal P}^4_\omega = {\cal P}^4_\omega - x{\cal P}^4_\lambda =
-\frac{4\epsilon_0^3 + 2s_2\epsilon_0 + s_1}{\xi^2}
+ O(\frac{1}{\xi}) =  O(\frac{1}{\xi})
\ee
Thus,
\be
d\lambda + xd\omega =
-\frac{\tilde{\cal P}^4_\omega}{{\cal P}^4_\lambda}d\omega
= O(1)d\xi,
\ee
as it should be ($d\lambda + xd\omega = d(\frac{1}{\xi} + \epsilon_0
+ \epsilon_1\xi + \ldots) + (\frac{1}{\xi^2} + O(\xi^2))d\xi =
(\epsilon_1 + O(1))d\xi$).

On the ``$-$'' sheet
\be
{\cal P}^4_\lambda = -\frac{64}{\xi^3} + O(\frac{1}{\xi}); \nn \\
{\cal P}^4_\omega = \frac{192}{\xi^5} + O(\frac{1}{\xi^3});\nn \\
\tilde{\cal P}^4_\omega = {\cal P}^4_\omega - x{\cal P}^4_\lambda =
\frac{256}{\xi^5} + O(\frac{1}{\xi^3}),
\ee
again in accordance with
\be
d\lambda + xd\omega = -\frac{\tilde{\cal P}^4_\omega}{{\cal P}^4_\lambda}
d\omega = \left(\frac{N_c}{\xi^2} + O(1)\right)d\xi =
 \left(\frac{4}{\xi^2} + O(1)\right)d\xi
\ee

\paragraph{Algebra of holomorphic differentials}

\be
dv_i = \frac{{\cal T}_id\omega}{{\cal P}^4_\lambda}, \ \ \
i = 0,1,2,3
\ee
We claim that for any
\be
dQ = \hat q_0dv_0 + \hat q_1dv_1 + \hat q_2 dv_2
\ee
and $i,j = 0,1,2$
\be
dv_idv_j = \left(\sum_{k=0,1,2} \hat C_{ij}^k dv_k\right) dQ
\ {\rm mod}\ (d\omega, d\lambda)
\label{algN4}
\ee

At the l.h.s. of (\ref{algN4}) there are some linear combination of
{\it six} linear independent quadratic combinations
${\cal T}_0^2$, ${\cal T}_0{\cal T}_1$,
${\cal T}_0{\cal T}_2$,  ${\cal T}_1^2$,
${\cal T}_1{\cal T}_2$ and ${\cal T}_2^2$, i.e.
$1$, $\lambda$, $\lambda^2-x$, $\lambda^2$, $\lambda(\lambda^2-x)$
and $(\lambda^2-x)^2$. At the r.h.s. we have:
\be
\left.
(\alpha_0{\cal T}_0 + \alpha_1{\cal T}_1 + \alpha_2{\cal T}_2)
(\hat q_0{\cal T}_0 + \hat q_1{\cal T}_1 + \hat q_2{\cal T}_2) +
(w_0{\cal T}_0 + w_1{\cal T}_1)\tilde P^3_\omega \ {\rm mod}\ P^3_\lambda
\ \right|_{{\cal P}^3 = 0}
\ee
If $w=0$, there is no possibility to adjust {\it three} parameters
$\alpha_0$, $\alpha_1$ and $\alpha_2$ to match {\it six} coefficients in
front of the relevant combinations ${\cal T}_i{\cal T}_j$.
$w_0$ and $w_1$ are additional {\it two} adjustment parameters,
and there is actually no need for the sixth one, since
${\cal T}_2^2$ is actually reducible to a linear combination
of the other five quadratic combinations.
Thus we should check that ${\cal P}^4_\omega = \tilde{\cal P}^4_\omega
\ {\rm mod}\ {\cal P}^4_\lambda$
and also ${\cal T}_2^2$ are equivalent to non-trivial
combination of ${\cal T}_0^2$, ${\cal T}_0{\cal T}_1$,
${\cal T}_0{\cal T}_2$, ${\cal T}_1^2$ and ${\cal T}_1{\cal T}_2$
(if ${\cal P}^4 = 0$ and {\it modulo} ${\cal P}^4_\lambda$).

First of all,
\be
\tilde{\cal P}^4_\omega =  -\lambda {\cal P}^4 +
{\cal T}_5 + s_2{\cal T}_3 + s_1{\cal T}_2 +
(s_0 + 12\mu_4 + 2g_2){\cal T}_1
\ee
Second,
\be
{\cal T}_4 = 4{\cal T}_1{\cal T}_3 - 3{\cal T}_2^2 - 3\mu_4; \nn \\
{\cal T}_5 = 3{\cal T}_1{\cal T}_4 - 2{\cal T}_2{\cal T}_3 -
6\mu_4{\cal T}_1; \nn \\
{\cal T}_1{\cal T}_5 = 4{\cal T}_2{\cal T}_4 - 3{\cal T}_3^2
+ 3\left(-(5\mu_4 + g_2){\cal T}_1^2 + (4\mu_4+g_2){\cal T}_2\right)
- 3g_3
\ee
Using these identities, one can get:
\be
\tilde {\cal P}^4_\omega \ {\rm mod}\ ({\cal P}^4, {\cal P}^4_\lambda)
= \nn \\ =
-2s_2 {\cal T}_1{\cal T}_2 - 3s_1{\cal T}_1^2 + \frac{3s_1}{2}{\cal T}_2
-(2s_0 + \frac{s_2^2}{2} - 6\mu_4 - 2g_2){\cal T}_1 - \frac{s_1s_2}{4}
\ee
and
\be
{\cal T}_1\tilde {\cal P}^4_\omega
\ {\rm mod}\ ({\cal P}^4, {\cal P}^4_\lambda)
= \nn \\ =
-3s_1 {\cal T}_1{\cal T}_2 +
(\frac{17s_2^2}{12} -3\mu_4 -g_2+s_0){\cal T}_1^2 + \nn \\
+(-\frac{4s_2^2}{3} - 4s_0 + 12\mu_4 + 3g_2) {\cal T}_2
- s_1s_2{\cal T}_1 -
(\frac{4s_0s_2}{3} - 4s_2\mu_4 + \frac{3s_1^2}{16} + 3g_3)
\ee
From these relations one can read off the structure constants of the
algebra. Since, the formulas are too long we are not going to presented them
here. The fact is, however, that the closed algebra exists, the structure
constants
were explicitly evaluated with the help of the program
MAPLE, and we used them to check that the algebra is indeed
{\it non}-associative, because
\be
[\hat C_0,\hat C_1]\sim q_2, \
\ \ [\hat C_0,\hat C_2]\sim q_1, \ \ \ [\hat C_1,\hat C_2]\sim q_0
\ee
(in general, $[C_i,C_j] \sim \oplus_{k\neq i,j} q_k$).

\section{Acknowledgments}

We are grateful to E.Akhmedov, M.Alishahiha, F.Ardalan,
J.De Boer, B.de Wit, B.Dubrovin, A.Gerasimov,
A.Gorsky, S.Gukov, J.Harnad, I.Krichever,
A.Losev, D.L\"ust, Yu.Manin, N.Nekrasov, M.Olshanetsky, A.Orlov, I.Polyubin,
A.Rosly, I.Tyutin and A.Zabrodin for useful discussions.

This work was supported by grants
RFBR-02-16117, INTAS-93-2058 (A.Mar.),
RFBR-96-02-16210(a), INTAS-96-2058 (A.Mir.) and RFBR 96-15-96939 (A.Mor.).
A.Mor. also acknowledges the support of DFG and the hospitality
of Humboldt University, Berlin, and IFH, Zeuthen.

\end{document}